\documentclass{IEEEtran}
\usepackage{amsfonts}
\usepackage{amssymb}
\usepackage{etex}
\usepackage[normalem]{ulem}
\usepackage[utf8]{inputenc}
\usepackage{url}
\usepackage{graphicx}
\usepackage{color}
\usepackage{array}
\usepackage{amsmath}
\usepackage{amsxtra}
\usepackage{array}
\usepackage{mathrsfs}
\usepackage{tikz}
\usetikzlibrary{arrows,calc,matrix,shapes,trees,positioning,automata}
\usepackage{algorithm}
\usepackage[noend]{algorithmic}
\usepackage{booktabs}
\usepackage{rotating}
\usepackage{paralist}
\usepackage{subfigure}
\usepackage{stmaryrd}
\usepackage{datetime}
\usepackage{epstopdf}
\usepackage{arydshln}
\usepackage[autolanguage]{numprint}
\usepackage{soul}
\usepackage[hidelinks]{hyperref}
\usepackage{centernot}
\usepackage{multirow}
\usepackage{tikz}
\usepackage{bm}
\usepackage{float}
\usepackage{mathtools}
\usepackage{mdwlist}
\usepackage{array}
\usepackage{enumitem}
\usepackage{cite}
\usepackage{caption}

\usepackage{amsthm}

\newcommand{\SWITCH}[1]{\STATE \textbf{switch} (#1) \begin{ALC@g}}
\newcommand{\ENDSWITCH}{\end{ALC@g}}
\newcommand{\CASE}[1]{\STATE \textbf{case} #1\textbf{:}}
\newcommand{\SIF}[1]{\STATE \textbf{if} #1}
\newcommand{\STHEN}[1]{\textbf{then} #1}

\newtheorem{example}{Example}
\newtheorem{theorem}{Theorem}

\newtheorem{lemma}{Lemma}

\newtheorem*{problem*}{Problem Statement}
\mathchardef\mhyphen="2D
\mathchardef\hh="2D
\newcommand\pp{\mathalpha{+}}

\newcommand\gap{\textrm{\textvisiblespace}\kern 0.5pt}

\newcommand\gen{\uparrow}
\newcommand\force{=}
\newcommand\Sup{\operatorname{Sup}}
\newcommand\Presup{\operatorname{Presup}}
\newcommand\anc{\operatorname{anc}}
\newcommand\desc{\operatorname{desc}}
\newcommand\Children{\operatorname{Children}}
\newcommand\List{\operatorname{Proj}}
\newcommand\PSup{\operatorname{Sup}}
\newcommand\Proj{\operatorname{Proj}}
\newcommand\ldot{.\!}

\newcommand\cc{\mathalpha{:}}

\newcommand\sra[1]{\!\xrightarrow{#1}\!}

\newcommand{\eat}[1]{} 

\newcommand{\sz}[1]{\lvert#1\rvert}   

\newcommand{\eqdef}{\stackrel{\mathrm{def}}{=}} 

\newcommand{\td}[2]{\if*#1\else^{#1}\fi\if*#2\else_{#2}\fi} 

\newcommand{\sset}[1]{\left\{\,#1\,\right\}} 

\newcommand\join\Join 

\DeclareSymbolFont{txsymbolsC}{U}{txsyc}{m}{n}
\SetSymbolFont{txsymbolsC}{bold}{U}{txsyc}{bx}{n}
\DeclareFontSubstitution{U}{txsyc}{m}{n}
\DeclareMathSymbol{\ljoin}{\mathrel}{txsymbolsC}{88}
\DeclareMathSymbol{\rjoin}{\mathrel}{txsymbolsC}{89}

\newsavebox\setminusbox
\newlength\setminuslen

\newcolumntype{C}{>{$\displaystyle}c<{$}} 
\newcolumntype{L}{>{$\displaystyle}l<{$}} 
\newcolumntype{R}{>{$\displaystyle}r<{$}} 

\newcommand{\B}[3]{B\if*#1\else_{#1}\fi(#2,#3)} 
\newcommand{\I}[3]{I\if*#1\else_{#1}\fi(#2,#3)} 

\makeatletter
\def\imod#1{\allowbreak\mkern10mu({\operator@font mod}\,\,#1)}
\makeatother

\newlength\hspaceoflen

\newcommand\vect[1]{{\boldsymbol{#1}}}
\newcommand\va{\vect{a}}
\newcommand\vb{\vect{b}}
\newcommand\vc{\vect{c}}
\newcommand\vd{\vect{d}}
\newcommand\ve{\vect{e}}
\newcommand\vf{\vect{f}}
\newcommand\vg{\vect{g}}
\newcommand\vh{\vect{h}}
\newcommand\vi{\vect{i}}
\newcommand\vj{\vect{j}}
\newcommand\vk{\vect{k}}
\newcommand\vl{\vect{l}}
\newcommand\vm{\vect{m}}
\newcommand\vn{\vect{n}}
\newcommand\vo{\vect{o}}
\newcommand\vp{\vect{p}}
\newcommand\vq{\vect{q}}
\newcommand\vr{\vect{r}}
\newcommand\vs{\vect{s}}
\newcommand\vt{\vect{t}}
\newcommand\vu{\vect{u}}
\newcommand\vv{\vect{v}}
\newcommand\vw{\vect{w}}
\newcommand\vx{\vect{x}}
\newcommand\vy{\vect{y}}
\newcommand\vz{\vect{z}}

\newcommand\mA{\vect{A}}
\newcommand\mB{\vect{B}}
\newcommand\mC{\vect{C}} 
\newcommand\mD{\vect{D}}
\newcommand\mE{\vect{E}}
\newcommand\mF{\vect{F}}
\newcommand\mG{\vect{G}}
\newcommand\mH{\vect{H}}
\newcommand\mI{\vect{I}}
\newcommand\mJ{\vect{J}}
\newcommand\mK{\vect{K}}
\newcommand\mL{\vect{L}}
\newcommand\mM{\vect{M}}
\newcommand\mN{\vect{N}} 
\newcommand\mO{\vect{O}}
\newcommand\mP{\vect{P}}
\newcommand\mQ{\vect{Q}} 
\newcommand\mR{\vect{R}} 
\newcommand\mS{\vect{S}}
\newcommand\mT{\vect{T}}
\newcommand\mU{\vect{U}}
\newcommand\mV{\vect{V}}
\newcommand\mW{\vect{W}}
\newcommand\mX{\vect{X}}
\newcommand\mY{\vect{Y}}
\newcommand\mZ{\vect{Z}}

\newcommand\bN{\mathbb{N}} 

\newcommand\xD{\mathscr{D}}

\DeclareMathAlphabet{\mathcal}{OMS}{cmsy}{m}{n}
\newcommand\cA{\mathcal{A}}

\newcommand\cD{\mathcal{D}}

\accentedsymbol\Ahat{{\hat A}}
\accentedsymbol\Bhat{{\hat B}}
\accentedsymbol\Chat{{\hat C}}
\accentedsymbol\Dhat{{\hat D}}
\accentedsymbol\Ehat{{\hat E}}
\accentedsymbol\Fhat{{\hat F}}
\accentedsymbol\Ghat{{\hat G}}
\accentedsymbol\Hhat{{\hat H}}
\accentedsymbol\Ihat{{\hat I}}
\accentedsymbol\Jhat{{\hat J}}
\accentedsymbol\Khat{{\hat K}}
\accentedsymbol\Lhat{{\hat L}}
\accentedsymbol\Mhat{{\hat M}}
\accentedsymbol\Nhat{{\hat N}}
\accentedsymbol\Ohat{{\hat O}}
\accentedsymbol\Phat{{\hat P}}
\accentedsymbol\Qhat{{\hat Q}}
\accentedsymbol\Rhat{{\hat R}}
\accentedsymbol\Shat{{\hat S}}
\accentedsymbol\That{{\hat T}}
\accentedsymbol\Uhat{{\hat U}}
\accentedsymbol\Vhat{{\hat V}}
\accentedsymbol\What{{\hat W}}
\accentedsymbol\Xhat{{\hat X}}
\accentedsymbol\Yhat{{\hat Y}}
\accentedsymbol\Zhat{{\hat Z}}

\accentedsymbol\ahat{{\hat a}}
\accentedsymbol\bhat{{\hat b}}
\accentedsymbol\chat{{\hat c}}
\accentedsymbol\dhat{{\hat d}}
\accentedsymbol\ehat{{\hat e}}
\accentedsymbol\fhat{{\hat f}}
\accentedsymbol\ghat{{\hat g}}
\accentedsymbol\hhat{{\hat h}}
\accentedsymbol\ihat{{\hat i}}
\accentedsymbol\jhat{{\hat j}}
\accentedsymbol\khat{{\hat k}}
\accentedsymbol\lhat{{\hat l}}
\accentedsymbol\mhat{{\hat m}}
\accentedsymbol\nhat{{\hat n}}
\accentedsymbol\ohat{{\hat o}}
\accentedsymbol\phat{{\hat p}}
\accentedsymbol\qhat{{\hat q}}
\accentedsymbol\rhat{{\hat r}}
\accentedsymbol\shat{{\hat s}}
\accentedsymbol\that{{\hat t}}
\accentedsymbol\uhat{{\hat u}}
\accentedsymbol\vhat{{\hat v}}
\accentedsymbol\what{{\hat w}}
\accentedsymbol\xhat{{\hat x}}
\accentedsymbol\yhat{{\hat y}}
\accentedsymbol\zhat{{\hat z}}

\accentedsymbol\rhohat{{\hat\rho}}

\accentedsymbol\Abar{{\bar A}}
\accentedsymbol\Bbar{{\bar B}}
\accentedsymbol\Cbar{{\bar C}}
\accentedsymbol\Dbar{{\bar D}}
\accentedsymbol\Ebar{{\bar E}}
\accentedsymbol\Fbar{{\bar F}}
\accentedsymbol\Gbar{{\bar G}}
\accentedsymbol\Hbar{{\bar H}}
\accentedsymbol\Ibar{{\bar I}}
\accentedsymbol\Jbar{{\bar J}}
\accentedsymbol\Kbar{{\bar K}}
\accentedsymbol\Lbar{{\bar L}}
\accentedsymbol\Mbar{{\bar M}}
\accentedsymbol\Nbar{{\bar N}}
\accentedsymbol\Obar{{\bar O}}
\accentedsymbol\Pbar{{\bar P}}
\accentedsymbol\Qbar{{\bar Q}}
\accentedsymbol\Rbar{{\bar R}}
\accentedsymbol\Sbar{{\bar S}}
\accentedsymbol\Tbar{{\bar T}}
\accentedsymbol\Ubar{{\bar U}}
\accentedsymbol\Vbar{{\bar V}}
\accentedsymbol\Wbar{{\bar W}}
\accentedsymbol\Xbar{{\bar X}}
\accentedsymbol\Ybar{{\bar Y}}
\accentedsymbol\Zbar{{\bar Z}}

\accentedsymbol\abar{{\bar a}}
\accentedsymbol\bbar{{\bar b}}
\accentedsymbol\cbar{{\bar c}}
\accentedsymbol\dbar{{\bar d}}
\accentedsymbol\ebar{{\bar e}}
\accentedsymbol\fbar{{\bar f}}
\accentedsymbol\gbar{{\bar g}}
\makeatletter
\@ifundefined{hbar}{}{
        \let\hbar\@undefined
}
\makeatother
\accentedsymbol\hbar{{\bar h}}
\accentedsymbol\ibar{{\bar i}}
\accentedsymbol\jbar{{\bar j}}
\accentedsymbol\kbar{{\bar k}}
\accentedsymbol\lbar{{\bar l}}
\accentedsymbol\mbar{{\bar m}}
\accentedsymbol\nbar{{\bar n}}
\makeatletter
\@ifundefined{obar}{}{
        \let\obar\@undefined      
}
\makeatother
\accentedsymbol{\obar}{{\bar o}}        
\accentedsymbol\pbar{{\bar p}}
\accentedsymbol\qbar{{\bar q}}
\accentedsymbol\rbar{{\bar r}}
\accentedsymbol\sbar{{\bar s}}
\accentedsymbol\tbar{{\bar t}}
\accentedsymbol\ubar{{\bar u}}
\accentedsymbol\vbar{{\bar v}}
\accentedsymbol\wbar{{\bar w}}
\accentedsymbol\xbar{{\bar x}}
\accentedsymbol\ybar{{\bar y}}
\accentedsymbol\zbar{{\bar z}}

\newcommand\eps{\epsilon}

\accentedsymbol\mAhat{{\hat\mA}}
\accentedsymbol\mBhat{{\hat\mB}}
\accentedsymbol\mChat{{\hat\mC}}
\accentedsymbol\mDhat{{\hat\mD}}
\accentedsymbol\mEhat{{\hat\mE}}
\accentedsymbol\mFhat{{\hat\mF}}
\accentedsymbol\mGhat{{\hat\mG}}
\accentedsymbol\mHhat{{\hat\mH}}
\accentedsymbol\mIhat{{\hat\mI}}
\accentedsymbol\mJhat{{\hat\mJ}}
\accentedsymbol\mKhat{{\hat\mK}}
\accentedsymbol\mLhat{{\hat\mL}}
\accentedsymbol\mMhat{{\hat\mM}}
\accentedsymbol\mNhat{{\hat\mN}}
\accentedsymbol\mOhat{{\hat\mO}}
\accentedsymbol\mPhat{{\hat\mP}}
\accentedsymbol\mQhat{{\hat\mQ}}
\accentedsymbol\mRhat{{\hat\mR}}
\accentedsymbol\mShat{{\hat\mS}}
\accentedsymbol\mThat{{\hat\mT}}
\accentedsymbol\mUhat{{\hat\mU}}
\accentedsymbol\mVhat{{\hat\mV}}
\accentedsymbol\mWhat{{\hat\mW}}
\accentedsymbol\mXhat{{\hat\mX}}
\accentedsymbol\mYhat{{\hat\mY}}
\accentedsymbol\mZhat{{\hat\mZ}}

\accentedsymbol\vahat{{\hat\va}}
\accentedsymbol\vbhat{{\hat\vb}}
\accentedsymbol\vchat{{\hat\vc}}
\accentedsymbol\vdhat{{\hat\vd}}
\accentedsymbol\vehat{{\hat\ve}}
\accentedsymbol\vfhat{{\hat\vf}}
\accentedsymbol\vghat{{\hat\vg}}
\accentedsymbol\vhhat{{\hat\vh}}
\accentedsymbol\vihat{{\hat\vi}}
\accentedsymbol\vjhat{{\hat\vj}}
\accentedsymbol\vkhat{{\hat\vk}}
\accentedsymbol\vlhat{{\hat\vl}}
\accentedsymbol\vmhat{{\hat\vm}}
\accentedsymbol\vnhat{{\hat\vn}}
\accentedsymbol\vohat{{\hat\vo}}
\accentedsymbol\vphat{{\hat\vp}}
\accentedsymbol\vqhat{{\hat\vq}}
\accentedsymbol\vrhat{{\hat\vr}}
\accentedsymbol\vshat{{\hat\vs}}
\accentedsymbol\vthat{{\hat\vt}}
\accentedsymbol\vuhat{{\hat\vu}}
\accentedsymbol\vvhat{{\hat\vv}}
\accentedsymbol\vwhat{{\hat\vw}}
\accentedsymbol\vxhat{{\hat\vx}}
\accentedsymbol\vyhat{{\hat\vy}}
\accentedsymbol\vzhat{{\hat\vz}}

\newcommand{\NAME}{DESQ}

\begin{document}
\title{DESQ: Frequent Sequence Mining \\ with Subsequence Constraints \\ {\large Technical Report, October 2016}}

\author{Kaustubh Beedkar \hspace{2cm} Rainer Gemulla \\
Data and Web Science Group\\
University of Mannheim, Germany\\
\{kbeedkar, rgemulla\}@uni-mannheim.de}

\maketitle  

\begin{abstract}
  Frequent sequence mining methods often make use of constraints to control
  which subsequences should be mined. A variety of such subsequence
  constraints has been studied in the literature, including length, gap, span,
  regular-expression, and hierarchy constraints. In this paper, we show that
  many subsequence constraints---including and beyond those considered in the
  literature---can be unified in a single framework. A unified treatment
  allows researchers to study jointly many types of subsequence constraints
  (instead of each one individually) and helps to improve usability of pattern
  mining systems for practitioners. In more detail, we propose a set of simple
  and intuitive ``pattern expressions'' to describe subsequence constraints
  and explore algorithms for efficiently mining frequent subsequences under
  such general constraints. Our algorithms translate pattern expressions to
  compressed finite state transducers, which we use as computational model,
  and simulate these transducers in a way suitable for frequent sequence
  mining. Our experimental study on real-world datasets indicates that our
  algorithms---although more general---are competitive to existing
  state-of-the-art algorithms.
\end{abstract}


\section{Introduction} 
\label{sec:introduction}

\emph{Frequent sequence mining} (FSM) is a fundamental task in data mining.
Frequent sequences are useful for a wide range of applications, including
market-basket analysis~\cite{gsp}, web usage mining and session
analysis~\cite{wum}, natural language processing~\cite{Lopez}, information
extraction~\cite{reverb, patty}, or computational biology~\cite{bio1}. In web
usage mining, for example, frequent sequences describe common behavior across
users (e.g., the order in which users visit web pages). As another example,
frequent textual patterns such as ``\emph{PERSON} is married to
\emph{PERSON}'' are indicative of typed relations between entities and useful
for natural-language processing and information extraction
tasks~\cite{reverb,patty}.

In FSM, we model the available data as a collection of sequences composed of
items such as words (text processing), products (market-basket analysis), or
actions and events (session analysis). Often items are arranged in an
application-specific hierarchy; e.g., \emph{is}$\to$\emph{be}$\to$\emph{VERB}
(for words), \emph{Canon 5D}$\to$\emph{DSLR camera}$\to$\emph{electronics}
(for products), or \emph{Rakesh
Agrawal}$\to$\emph{scientist}$\to$\emph{PERSON} (for entities). The goal of
FSM is to discover subsequences or generalized subsequences that occur in
sufficiently many input sequences.  Since the total number of such
subsequences can potentially be very large and not all frequent subsequences
may be of interest to a particular application, most FSM methods make use of
subsequence constraints to control the set of subsequences to be mined.

A large variety of subsequence constraints has been studied in prior
work~\cite{ctg, lash, spirit, mgfsm, prefixspan, prefix_monotone, gsp,
cspade}.  Commonly proposed constraints include \emph{gap or span
constraints}, where items in the subsequences need to appear ``close'' in the
input sequence, and \emph{length constraints}, where the number of items in
the subsequences is bounded. In $n$-gram mining~\cite{Berberich-ngrams}, for
example, the goal is to mine frequent consecutive subsequences of exactly $n$
words. \emph{Hierarchy constraints} allow controlled generalization according
to the item hierarchy to find patterns which do not directly occur in the
input data. Examples include shopping patterns such as ``customers frequently
buy some \emph{DSLR camera}, then some \emph{tripod}, then some \emph{flash}'' or
textual patterns such as ``\emph{PERSON} \emph{be} born in \emph{LOCATION}''.
\emph{Regular expression (RE) constraints} have also been studied in the
context of FSM; here subsequences must match a given RE.

A number of specialized algorithms for various combinations of the above
subsequence constraints have been proposed in the literature. In this paper,
we show that many subsequence constraints---including and beyond those
described above---can be unified in a single framework. A unified treatment
allows researchers to study subsequence constraints in general instead of
focusing on certain combinations individually. It also helps to improve
usability of pattern mining systems for practitioners because it avoids the
need to develop customized mining algorithms for the particular subsequence
constraint of interest. In this work, we focus on the questions of (1) how to
model and express subsequence constraints in a suitable way and (2) how to
mine efficiently all frequent sequences that satisfy the given constraints.

In more detail, we introduce \emph{subsequence predicates} to model subsequence
constraints in a general way, and we propose a simple and intuitive
\emph{pattern expression language} to concisely express subsequence
predicates. Our pattern expressions are based on regular expressions, but---in
contrast to prior work on RE-constrained FSM---target input sequences and
support capture groups and item hierarchies. Capture groups are the key
ingredient for expressing most prior subsequence constraints in a unified way;
see Tab.~\ref{tab:existing_relevance} for examples. Direct support for item
hierarchies allows us both to express subsequence constraints concisely and to
mine generalized subsequences in a controlled way. Some example pattern
expressions as well as anecdotal results are given in
Tab.~\ref{xtab:exp_pattern_expressions}.

To mine frequent sequences, we propose to use finite state transducers (FST)
as the underlying computational model. To the best of our knowledge, FSTs have
not been studied in the context of FSM before. We propose the \NAME{} system,
which includes two efficient mining algorithms termed \NAME-COUNT and
\NAME-DFS. Both algorithms translate a given pattern expression to a
\emph{compressed} FST, which is subsequently optimized and simulated in a way
suitable for frequent sequence mining. \NAME-COUNT is a match-and-count
algorithm that aims at highly selective constraints, whereas \NAME-DFS can
handle more demanding pattern expressions and is inspired by
PrefixSpan~\cite{prefixspan}. Our experimental study on various real-world
datasets suggests that \NAME{} is an efficient general-purpose FSM framework
and competitive to state-of-the-art specialized algorithms.


\section{Preliminaries}

\textbf{Sequence database.} A \emph{sequence database} is a multiset of
sequences, denoted $\xD = \sset{T_1,T_2,\ldots,T_{\sz{\xD}}}$.  Each
\emph{sequence} $T=t_1t_2\ldots t_{\sz{T}}$ is an ordered list of items from a
\emph{vocabulary} $\Sigma=\sset{w_1,w_2,\ldots,w_{\sz{\Sigma}}}$. We denote by
$\eps$ the empty sequence, by $\sz{T}$ the length of sequence $T$, by
$\Sigma^*$ ($\Sigma^+$) the set of all (all non-empty) sequences that can be
constructed from items in $\Sigma$. Fig.~\ref{fig:example_database} shows an
example sequence database $\xD_{ex}$ consisting of six sequences.

\textbf{Item hierarchy.} The items in $\Sigma$ are arranged in an
\emph{item hierarchy}, which expresses how items can be generalized (or that
they cannot be generalized). Fig.~\ref{fig:example_hierarchy} shows an example
hierarchy in which, for example, item $a_1$ generalizes to item $A$. In
general, we say that an item $u$
\emph{directly generalizes to} an item $v$, denoted $u \Rightarrow v$, if $u$ is
a child of $v$. We further denote by $\Rightarrow^*$ the reflexive transitive
closure of $\Rightarrow$. For the example of Fig.~\ref{fig:example_hierarchy},
we have $b_{11} \Rightarrow b_1$, $b_1 \Rightarrow B$ and $b_{11}
\Rightarrow^* B$. For each item $w\in\Sigma$, we denote by \[\anc(w) =
\sset{w'\mid w
\Rightarrow ^* w'} \] the set of \emph{ancestors} of $w$ (including $w$) and by
\[ \desc(w) = \sset{w' \mid w' \Rightarrow^* w} \] the set of \emph{descendants}
of $w$ (again, including $w$). In our running example, we have $\anc(b_1) =
\sset{b_1,B}$ and $\desc(b_1)=\sset{b_1,b_{11},b_{12}}$.

\usetikzlibrary{positioning}
\tikzstyle{block}=[draw,
fill=white,
rounded corners, 
minimum width=4mm,
minimum height=4mm,
inner sep=1pt]
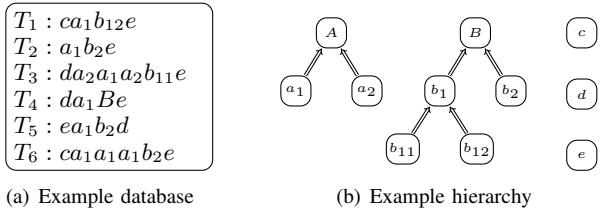
\begin{figure}


\centering
\subfigure[Example database]{
\label{fig:example_database}
\vspace{0pt}
\hspace{.4cm}
\begin{tikzpicture}\small
\node[draw, rounded corners] {
\begin{minipage}{0.14\textwidth}
  $T_1:ca_1b_{12}e$\\
  $T_2:a_1b_2e$ \\
  $T_3:da_2a_1a_2b_{11}e$\\
  $T_4:da_1Be$ \\
  $T_5:ea_1b_2d$\\
  $T_6:ca_1a_1a_1b_2e$
\end{minipage}
};
\end{tikzpicture}\hspace{.4cm}
}
\subfigure[Example hierarchy]{
\label{fig:example_hierarchy}
\vspace{0pt}
\begin{tikzpicture}
\tiny
\node[style=block] (A) at (0,0) {$A$};
\node[below=5mm of A] (dummyA) {};
\node[style=block, left=2mm of dummyA] (a1) {$a_1$};
\node[style=block, right=2mm of dummyA] (a2) {$a_2$};

\node[style=block, right=15mm of A] (B) {$B$};
\node[below=5mm of B] (dummyB) {};
\node[style=block, left=2mm of dummyB] (b1) {$b_1$};
\node[style=block, right=2mm of dummyB] (b2) {$b_2$};

\node[below=5mm of b1] (dummyb1) {};
\node[style=block, left=2mm of dummyb1] (b11) {$b_{11}$};
\node[style=block, right=2mm of dummyb1] (b12) {$b_{12}$};

\node[style=block, right=10mm of B] (c) {$c$};
\node[style=block, below=4mm of c] (d) {$d$};
\node[style=block, below=4mm of d] (e) {$e$};

\draw[-implies,double] (a1) -- (A);
\draw[-implies,double] (a2) -- (A);
\draw[-implies,double] (b1) -- (B);
\draw[-implies,double] (b2) -- (B);
\draw[-implies,double] (b11) -- (b1);
\draw[-implies,double] (b12) -- (b1);

\end{tikzpicture}
}
\caption{A sequence database and its vocabulary}
\label{fig:example_database_hierarchy}
\end{figure}


\textbf{Subsequence.} Let $S = s_1s_2\ldots s_{\sz{S}}$ and $T =
t_1t_2\ldots t_{\sz{T}}$ be two sequences composed of items from $\Sigma$. We
say that $S$ is a \emph{generalized subsequence} of $T$, denoted $S
\sqsubseteq T$, if $S$ can be obtained by deleting and/or generalizing items
in $T$. More formally, $S
\sqsubseteq T$ iff there exists integers $1 \le i_1 < i_2 < \dots < i_{\sz{S}}
\le \sz{T}$ such that $t_{i_k} \Rightarrow^* s_k$ for $1\le k\le
\sz{S}$. Continuing our example, we have $cBe \sqsubseteq T_1$, $ca_1
\sqsubseteq T_1$ and $a_1c \not\sqsubseteq T_1$. 
\eat{As described above, the use of
generalized subsequences allows for mining generalized sequential patterns,
which may not directly occur in the database.} \eat{In the context of text
mining, for example, such patterns may include generalized $n$-grams (``the
\emph{ADJ} house'') or typed relational patterns (``\emph{PERSON} lives in
\emph{CITY}'').  }


\section{FSM With Subsequence Constraints}

\textbf{Subsequence constraints.}  A subsequence constraint describes which
subsequences of a given input sequence should be considered for frequent
sequence mining. Our goal is to provide a general framework to express
subsequence constraints, including and beyond previously proposed constraints.
Consider the following (admittedly contrived) subsequence constraint as an
example.

\begin{example}\label{ex:psubseq1} Consider our example database $\xD_{ex}$
and suppose that we are interested in mining sequences of $B$'s and/or
descendants of $A$'s. We restrict attention to sequences that occur
consecutively in input sequences starting with $c$ or $d$ and ending with
$e$. We also allow to generalize occurrences of descendants of $A$ and $B$.
Then $a_1B\sqsubseteq T_1$ and $AB\sqsubseteq T_1$ satisfy this subsequence
constraint, whereas $a_1b_{12}\sqsubseteq T_1$, $a_1b_1\sqsubseteq T_1$,
$a_1B\sqsubseteq T_2$ and $AB\sqsubseteq T_2$ do not. 
\end{example}

The above subsequence constraint cannot be expressed using prior methods. Note
that the constraint combines (i) a gap constraint (consecutive), (ii) a
hierarchy constraint (descendants of $B$ must be generalized), and (iii) a
context constraint (between $c$ or $d$, and $e$).

\textbf{Subsequence predicates.} We propose subsequence predicates as
a general, natural model for subsequence constraints. A \emph{subsequence
predicate} $P$ is a predicate on pairs $(S,T)$, where $T\in\Sigma^+$ is any
input sequence and $S\sqsubseteq T$ is a subsequence. Subsequence
$S\sqsubseteq T$ satisfies the constraint when $P(S,T)$ holds. Note that $P$
is not a predicate on (only) subsequence $S$; it also involves input sequence
$T$. We denote by \[ G_P(T)=\sset{S\sqsubseteq T\mid P(S,T)} \] the set of
\emph{$P$-subsequences} in $T$. For each $S\in G_P(T)$, we say that $S$ is
\emph{$P$-generated} by $T$. For example, let $P_{ex}$ be the subsequence
predicate that expresses subsequence constraint of Ex.~\ref{ex:psubseq1}, then
$G_{P_{ex}}(T_1)=\{a_1B, AB\}$ and $G_{P_{ex}}(T_2)=\emptyset$.

Subsequence predicates can encode different application needs, including but
not limited to the various subsequence constraints discussed before.  A
subsequence predicate can act as a filter on the set of all subsequences of
$T$ (only $A$'s and $B$'s), but may also consider the context in which these
subsequences occur (consecutively between $c$ or $d$ and $e$). In practice, we
may construct subsequence predicates that generate all $n$-grams, all
adjective-noun pairs, all relational phrases between named entities, all
electronic products, or, in log mining, sequences of items that occur before
and/or after an error item.  We propose a suitable way to express subsequence
predicates in Sec.~\ref{sec:pattern_expressions}.

\textbf{FSM and subsequence predicates.} Let $P$ be a subsequence predicate. The
\emph{$P$-support} $\Sup_P(S,\xD)$ of sequence $S\in\Sigma^+$ in sequence
database $\xD$ is the \emph{multiset} of all sequences in $\xD$ that
$P$-generate $S$, i.e.,
\begin{equation}
\Sup_P(S,\xD) = \sset{T \in \xD \mid S \in G_P(T)}. \label{eq:sup}
\end{equation}

The \emph{$P$-frequency} of $S$ in $\xD$ is given by
\[ f_P(S,\xD)=\sz{\Sup_P(S,\xD)}.\] In our example database, we have
$\Sup_{P_{ex}}(Aa_1AB,\xD_{ex})=\sset{T_3,T_6}$ and thus
$f_{P_{ex}}(Aa_1AB,\xD_{ex})=2$. Given a \emph{support threshold} $\sigma>0$,
we say that a sequence $S$ is \emph{$P$-frequent} if $f_P(S,\xD)\ge
\sigma$.

\begin{problem*}
  Given a sequence database $\xD$, a subsequence predicate $P$, and a support
  threshold $\sigma>0$, find all $P$-frequent sequences $S\in\Sigma^+$ along
  with their frequencies.
\end{problem*}

The set of all $P_{ex}$-frequent sequences for $\sigma=2$ in our example
database is given by
\[
\sset{AAAB \cc 2, AB \cc 2, Aa_1AB \cc 2, a_1B \cc 2},
\]
where we also give $P$-frequencies.

\textbf{Discussions.} The above definitions are generalizations of the
notions of frequency and support used in traditional frequent sequence mining.
Efficient mining of $P$-frequent sequences is challenging because the
antimonotonicity property does not hold directly: We cannot generally deduce
from the knowledge that sequence $S$ is $P$-frequent whether or not any of the
subsequences of $S$ are $P$-frequent as well. Nevertheless, our mining
algorithms make use of suitable adapted notions of antimonotonicity for
subsequence predicates (Lemma~\ref{lemma:item_monotonicity}) and pattern
expressions (Lemma~\ref{lem:prefix_monotonicity}).

\section{Pattern Expressions} \label{sec:pattern_expressions}

We propose a pattern language for expressing subsequence predicates in a
simple and intuitive way. Our language is based on regular expressions, but
adds features that allows us to unify many prior subsequence constraints. We
subsequently suggest a computational model based on FSTs, and describe the
formal semantics of our language.

\subsection{Pattern Language}

Our language consists of the following set of \emph{pattern expressions},
defined inductively:
\begin{enumerate}
\item For each item $w\in\Sigma$, the expressions $w$, $w_\force$, $w^\gen$, and
$w_\force^\gen$ are pattern expressions. 
\item $.$ and $.^\gen$ are pattern expressions.
\item If $E$ is a pattern expression, so are $(E)$, $[E]$, $[E]^*$, $[E]^+$,
$[E]?$, and for all $n,m\in \bN$ with $n\le m$, $[E]\{n\}$, $[E]\{n,\}$, and
$[E]\{n,m\}$.
\item If $E_1$ and $E_2$ are pattern expressions, so are $[E_1E_2]$ and
$[E_1|E_2]$.
\end{enumerate}

Pattern expressions are based on regular expressions, but additionally include
capture groups (in parentheses), hierarchies (by omitting $_=$), and
generalizations (using $^\gen$ and $_\force^\gen$). We make use of the usual
precedence of rules for regular expressions to suppress square brackets (but
not parentheses); operators that appear earlier in the above definition have
higher precedence. We refer to expressions of form (1) or (2) as \emph{item
expressions}. We write $G_E(T)$ to refer to the set of subsequences
``generated'' by expression $E$ on input $T$ (see
Sec.~\ref{sec:pattern_matching} for a formal definition).

\textbf{Captured and uncaptured expressions.}  Pattern expressions specify which
subsequences to output (captured) as well as the context in which these
subsequences should occur (uncaptured). We make use of parentheses to
distinguish these two cases; the semantics is similar to the use of capture
groups in regular expressions. Given an expression $E$, only subexpressions
that are enclosed in or contain a capture group will produce non-empty output;
all other subexpressions serve to describe context information. For example,
the pattern expression

\begin{equation}\label{eq:patex}
E_{ex}=[c|d]([A^\gen\mid B_\force^\gen]^+)e
\end{equation}
describes precisely the subsequence constraint of Ex.~\ref{ex:psubseq1}. Here
subexpressions $[c|d]$ and $e$ describe context and $([A^\gen\mid
B_\force^\gen]^+)$ output.

\textbf{Item expressions.} Item expressions are the elementary form of pattern
expressions and apply to one input item. If the item expression ``matches'' the
input item, it can ``produce'' an output item; see
Tab.~\ref{tab:item_expressions} for an overview. Fix some $w\in\Sigma$. The most
basic item expression is $w_\force$: it matches only item $w$ and produces
either $\eps$ (if uncaptured) or $w$ (if captured). Using our example hierarchy
of Fig.~\ref{fig:example_fst}, we have $G_{A_\force}(A) = \emptyset$ (note
that we ignore output $\eps$), $G_{(A_\force)}(A) = \sset{A}$, and
$G_{(A_\force)}(a_1) = \emptyset$. Sometimes we do not want to only match the
specified item but also all of its descendants in the item hierarchy (e.g., we
want to match all nouns in text mining). Item expression $w$ serves this
purpose: it matches any item $w'\in\desc(w)$ (which includes $w$) and, when
captured, produces the item that has been matched. For example, we have
$G_{(A)}(A) = \sset{A}$, $G_{(A)}(a_1) = \sset{a_1}$, and $G_{(A)}(b_1) =
\emptyset$. Our language also provides \emph{wild card} symbol ``.'' to match
any item; again, the matched item is produced when the wild card is captured.
For example, $G_{(.)}(A) = \sset{A}$, and $G_{(.)}(a_1) = \sset{a_1}$.

To support mining with controlled generalizations (e.g., to mine patterns such as
``\emph{PERSON} lives in \emph{CITY}''), we use the \emph{generalization
  operator} $^\gen$, which generalizes items along the hierarchy. Item
expressions that use the generalization operator must be captured. More
specifically, item expression $w^\gen$ matches any item $w'\in\desc(w)$---as
expression $w$ does\mbox{---,} and it produces either the matched input item or any of
its ancestors that is also a descendant of $w$. For example,
$G_{(B^\gen)}(b_{12}) = \sset{b_{12}, b_1, B}$ and $G_{(b_1^\gen)}(b_{12}) =
\sset{b_{12}, b_1}$. We also allow the use of a wild card with generalization
operator: expression ``$.^\gen$'' matches any item and produces each of its
generalizations. For example, $G_{(.^\gen)}(b_1) = \sset{b_1,B}$. Our final
item expression is used to enforce a generalization: $w_\force^\gen$ matches
any descendant of $w$ and produces $w$, independently of which descendant has
been matched. For example $G_{(B_\force^\gen)}(b_{12}) = \sset{B}$.

\begin{table}[t]
\newcolumntype{R}{@{\extracolsep{-110pt}}>{\it}r@{\extracolsep{-28pt}}}%
\centering
\caption{Pattern expr. for prior subsequence constraints}
\begin{tabular}{lRr}
  \toprule
  Subsequence constraint & Example & Pattern expression \\
  \midrule
  All subsequences~\cite{gsp, spade, prefixspan} & & $[.^*(.)]^+$ \\
  Bounded length~\cite{cspade} & length 3--5  & $[.^*(.)]\{3,5\}$ \\
  $n$-grams~\cite{mgfsm,Berberich-ngrams} & $3$-, $4$- and $5$-grams & $(.)\{3,5\}$ \\
  Bounded gap~\cite{cspade,mgfsm} & each gap at most 3 & $(.)[.\{0,3\}(.)]+$ \\
  Serial episodes~\cite{Mannila:1997fk} 
  & length 3, total gap $\le 2$ \\ && $(.)[.?.?(.)\mid$ $.?(.).?\mid(.).?.?](.)$\\
  Hierarchy~\cite{gsp,lash} & generalized 5-grams & $(.^\gen)\{5\}$\\
  Regular expression~\cite{spirit,hackle_tree,petri_net,prefix_monotone} & \\
  & subsequences matching $[a|b]\,c^*d$ & $(a|b)[.^*(c)]^*.^*(d)$ \\
  & contiguous subsequences matching $[a|b]\,c^*d$ & $([a|b]\,c^*d)$\\
  \bottomrule
\end{tabular}
\label{tab:existing_relevance}
\end{table}

\textbf{Composite expressions.} Item expressions can be arbitrarily combined
using operators $?$ (optionality), $*$ (Kleene star), $+$ (Kleene plus),
$\{n,m\}$ (bounded repetition), $|$ (union), and concatenation to match
(sequences of) more than one input item. The semantics of these compositions
is as in regular expressions.

\textbf{Examples.} Our pattern expressions allow us to express many
existing subsequence constraints in a unified way; see
Tab.~\ref{tab:existing_relevance} for some examples. Note that the use of
capture groups enables many of these pattern expressions. Pattern expressions
can additionally express many customized subsequence constraints that cannot
be handled by existing FSM frameworks; see
Tab.~\ref{xtab:exp_pattern_expressions} for some examples.


\begin{table*}[t]
\centering
\caption{Translation rules for item expressions (where $w,w',w''\in\Sigma$)}
\renewcommand\arraystretch{0.9}
\begin{tabular}{llllll} %
\toprule
Expr. & Matches & Transl.~type & Produces & FST & Compressed FST \\
\midrule
$w_\force$ & $w$ & Uncaptured & $\epsilon$ & $\{q_S\xrightarrow{w\cc \epsilon}q_F\}$ & $\{q_S\xrightarrow{w_\force\cc \epsilon}q_F\}$ \\
&  & Captured &  $w$ & $\{q_S\xrightarrow{w\cc w}q_F\}$ & $\{q_S\xrightarrow{w_\force\cc w}q_F\}$  \\

$w$ & $w'\in \desc(w)$ & Uncaptured & $\epsilon$ & $\{q_S\xrightarrow{w'\cc \epsilon}q_F \mid w'\in \desc(w)\}$ & $\{q_S\xrightarrow{w\cc \epsilon}q_F\}$ \\
&  & Captured &  $w'$ & $\{q_S\xrightarrow{w'\cc w'}q_F \mid w'\in \desc(w)\}$ & $\{q_S\xrightarrow{w\cc \$}q_F\}$  \\

$.$ & $w \in \Sigma$ & Uncaptured & $\epsilon$ & $\{q_S\xrightarrow{w\cc \epsilon} q_F\mid w \in \Sigma\}$ & $\{q_S\xrightarrow{.\cc \epsilon} q_F\}$\\
					 &									   & Captured   & $w$ & $\{q_S\xrightarrow{w\cc w} q_F\mid w \in \Sigma\}$ & $\{q_S \xrightarrow{.\cc \$} q_F\}$ \\

$w^\gen$ & $w'\in \desc(w)$ & Captured & $\anc(w') \cap \desc(w)$ & $\{q_S\xrightarrow{w'\cc w''}q_F \mid w' \in \desc(w), w'' \in \anc(w')\cap \desc(w)\}$   & $\{q_S\xrightarrow{w\cc \$\mhyphen w}q_F\}$ \\ 

$.^\gen$ & $w\in \Sigma$ & Captured & $\anc(w)$ &  $\{q_S \xrightarrow{w\cc w'} q_F \mid w \in \Sigma, w' \in \anc(w)\}$ & $\{q_S\xrightarrow{.\cc \$\mhyphen\top}q_F\}$\\

$w_\force^\gen$ & $w'\in \desc(w)$ & Captured & $w$ & $\{q_S\xrightarrow{w'\cc w}q_F \mid w' \in \desc(w)\}$ & $\{q_S\xrightarrow{w\cc w}q_F\}$ \\
\bottomrule
\end{tabular}
\label{tab:item_expressions}
\end{table*}


\subsection{Computational Model}\label{sec:pattern_matching}

We translate patterns expressions into FSTs, which are a natural computational
model for pattern expressions. An FST is a type of finite state machine for
string-to-string translation~\cite{Mohri:1997}. FSTs are similar to finite
state automata but additionally label transitions with output strings.
Conceptually, an FST reads an input string and translates it to an output
string in a nondeterministic fashion. We will use FSTs to specify subsequence
predicate $P(S,T)$: the predicate holds if the FST can output subsequence $S$
when reading input $T$.

\textbf{Finite state transducers.} More formally, we consider a restricted form
of FSTs defined as follows. An FST $\cA$ is a 5-tuple ($Q$, $q_S$, $Q_F$,
$\Sigma$, $\Delta$), where 
\begin{itemize}
\item $Q$ is a set of states,
\item  $q_S \in Q$ is the initial state,
\item $Q_F \subseteq Q$ is the set of final states,
\item $\Sigma$ is an input and output alphabet, and
\item $\Delta\subseteq Q\times(\Sigma\cup\sset{\eps})\times(\Sigma\cup\sset{\eps})\times Q$ is a transition relation.
\end{itemize}
    For every
transition $(q_{from},in,out,q_{to})\in\Delta$, we require that
$out\in\anc(in)\cup\sset{\eps}$ and that whenever $in=\eps$ then $out=\eps$.
Our notion of FSTs differs from traditional FSTs in that we use a common input
and output alphabet and in that we restrict output labels. The latter
restriction ensures that our FSTs output generalized subsequences of their
input (Lemma~\ref{lem:subsequence_program}). Fig.~\ref{fig:example_fst} shows
an example FST, where $q_S = q_0$, $Q_F = \sset{q_{11}}$, and each transition
is marked with $in\cc out$ labels. We refer to transitions with $in = \eps$
(and thus $out=\eps$) as \emph{$\eps$-transitions}; these transitions are
marked with $\eps$ in the figure.

\textbf{Runs and outputs.} 
Let $T=t_1t_2\ldots t_n$ be an input sequence. A \emph{run} for $T$ is a sequence
$p=p_1p_2\ldots p_{m}$ of transitions, where for $1\le i\le m$:
$p_i=(q_i,w_i,w_i',q_i')\in\Delta$, $q_1=q_S$, $q_{i+1}=q_i'$, and $w_1w_2\ldots
w_m=T$ (recall that $w_i\in\Sigma\cup\sset{\eps}$ so that $m\ge
n$). Intuitively, the FST starts in state $q_S$ and repeatedly selects
transitions that are consistent with the next input item. If $q_m\in Q_F$, we
refer to $p$ as an \emph{accepting run}. The output $O(p)$ of run $p$ is the
sequence $S=w_1'\ldots w_m'$ of output labels, where we omit all $w_i'$ with
$w_i'=\eps$ and set $S=\eps$ if all $w_i'=\eps$. The set of sequences
generated by FST $\cA$ is given by
\[
G_\cA(T) = \sset{ O(p)\neq\eps \mid \text{$p$ is an accepting run of $\cA$ for $T$}}.
\]

\begin{example}\label{ex:fst_run}
Consider the FST $\cA_{F\ref{fig:example_fst}}$ of Fig.~\ref{fig:example_fst}.
$\cA_{F\ref{fig:example_fst}}$ has two accepting runs for sequence
$T_1=ca_1b_{12}e$, which are given by
$p_1=q_0\sra{\eps}q_1\sra{c\cc\eps}q_3\sra{\eps}q_5\sra{\eps}q_6\sra{a_1\cc
a_1}q_8\sra{\eps}q_{10}\sra{\eps}q_5\sra{\eps}q_7\sra{b_{12}\cc
B}q_9\sra{\eps}q_{10}\sra{e\cc\eps} q_{11}$ with output $O(p_1)=a_1B$, and
$p_2=q_0\sra{\eps}q_1\sra{c\cc\eps}q_3\sra{\eps}q_5\sra{\eps}q_6\sra{a_1\cc
A}q_8\sra{\eps}q_{10}\sra{\eps}q_5\sra{\eps}q_7\sra{b_{12}\cc
B}q_9\sra{\eps}q_{10}\sra{e\cc\eps} q_{11}$ with output $O(p_2)=AB$. Thus,
$G_{\cA_{F\ref{fig:example_fst}}}(T_1)=\sset{a_1B,AB}$, as desired.  There is
no accepting run for $T_2$ so that
$G_{\cA_{F\ref{fig:example_fst}}}(T_2)=\emptyset$. Observe that
$\cA_{F\ref{fig:example_fst}}$ generates precisely the $P$-sequences of
Ex.~\ref{ex:psubseq1}.
\end{example}

The following lemma states that our FSTs generate generalized subsequences of
their inputs and thus specify subsequence predicates. Note that the lemma holds
for any run, whether or not accepting. 

\begin{lemma}\label{lem:subsequence_program}
  Let $T\in\Sigma^*$ be an input sequence and $\cA$ be an FST. For any run
  $p$ of $\cA$ for $T$, it holds $O(p)\sqsubseteq T$.
\end{lemma}

\begin{IEEEproof}
The proof is by induction. For $T=\eps$, the assertion holds because every
path for $T$ must consist of only $\eps$-transitions so that
$G(p)=\eps\sqsubseteq T$. Now suppose that the assertion holds for some
sequence $T'\in\Sigma^*$. We show that it then also holds for $T=T'w$,
$w\in\Sigma$. Let $p$ be any path for $T$ and set $S=O(p)$. We decompose $p$
into two sequences of transitions: a path $p'$ for $T'$ with output $S'$ and a
remainder $p_w$ with output $s_w$. This decomposition is always possible. We
have $S=S's_w$. Since $p'$ is a path for $T'$, $S'\sqsubseteq T'$
by the induction hypothesis. Now observe that $p_w$ must contain exactly one
transition with input label $w$ and that all other transitions must be
$\eps$-transitions; otherwise $p$ would not be a path for $T$. Let $w'$ be the
output label of the transition with input label $w$. Then $s_w=w'$. By the
definition of FSTs, we must have $w'\in\anc(w)\cup\sset{\eps}$, which implies
that $w'\sqsubseteq w$. Since $S'\sqsubseteq T'$ and $s_w\sqsubseteq w$, we
obtain $S=S's_w\sqsubseteq T'w=T$.
\end{IEEEproof}

Note that not all subsequence predicates can be expressed with FSTs; e.g.,
there is no FST for predicate ``all subsequences of form $a^*b^*$ with an
equal number of $a$'s and $b$'s''. FST are a good trade-off between
expressiveness and computational complexity, however: they can express many
subsequence constraints that occur in practice and they lend themselves to
efficient mining (see Sec.~\ref{sec:pattern_mining}).

\textbf{Translating pattern expression.} We now describe how to translate a
pattern expression $E$ into an FST $\cA(E)$. The FST formally defines the
semantics of pattern expressions: we set $G_E(T) \eqdef G_{\cA(E)}(T)$. Each
item expression is translated into a two-state FST with $Q=\sset{q_S,q_F}$,
where $q_S$ is the initial and $q_F$ the final state. The transitions of the
FST depend on the item expression and are summarized in
Tab.~\ref{tab:item_expressions}, column ``FST''. The translation rules for
composite expressions mirror the Thompson construction~\cite{Thompson68} for
translating regular expressions to finite state automata.\footnote{All
translation rules can be implemented without introducing any
$\eps$-transitions; we follow this approach in our actual implementation but
use $\eps$-transitions in our example FSTs for improved readability.} For
example, expression $E_{ex}$ of Eq.~\eqref{eq:patex} translates to the FST of
Fig.~\ref{fig:example_fst}.

\textbf{Compressed FST.} The translation rules above can produce very large
FSTs, especially when the vocabulary is large. For example, if the hierarchy
has $n$ items and average depth $d$, the FST for ``$.^\gen$'' has $\Theta(nd)$
transitions. To avoid this explosion of FST size and support efficient mining,
we make use of a compressed FST (cFST) representation for this purpose; see
column ``compressed FST'' of Tab.~\ref{tab:item_expressions}. The cFST of an
item expression has exactly one transition, but input and output labels are
taken from an alphabet larger than $\Sigma$. Each transition in the cFST
describes a set of transitions in the corresponding FST in a concise way. More
specifically, cFSTs use as input labels $.$, $w$, and $w_=$ for all
$w\in\Sigma$. Here ``.'' matches all input items, $w$ matches all items in
$\desc(w)$, and $w_=$ matches only item $w$.  cFSTs use as output labels
$\eps$, $w$, $\$$, $\$\hh w$, and $\$\hh\top$ for $w\in\Sigma$. Each
transition encodes the set of output labels in the corresponding FST: $\eps$
and $w$ are as before, $\$$ encodes the matched input item, $\$\hh w$ the
matched input item and all its ancestors up to $w$, and $\$\hh\top$ the
matched item and all its ancestors. The cFST translations for composite
expressions remain unmodified. Fig.~\ref{fig:example_compressed_fst} shows a
cFST $\cA_{ex}$ for $E_{ex}$ (Eq.~\eqref{eq:patex}). Note that the cFST has
fewer transitions than its uncompressed counterpart of
Fig.~\ref{fig:example_fst}.

\textbf{Simulating cFSTs.}
A simple way to compute the set $G_\cA(T)$ for a cFST
is via simulation and backtracking. Note that the computation of $G_\cA(T)$ for
all $T\in \cD$ can be infeasible. Nevertheless, simulation forms the basis of
the more efficient \NAME-DFS algorithm of Sec.~\ref{subsec:p-incr} so that we
describe the approach briefly. Denote by the \emph{transition function}
\[
\delta(q,w) = \lbrace\,(out,q_{to}) \mid (q,in,out,q_{to}) \in 
\Delta,\,in \text{ matches } w \,\rbrace
\]
the set of (output label, state)-pairs that can be reached from state $q$ by
consuming input item $w$ (see column ``Matches'' in
Tab.~\ref{tab:item_expressions}). To simulate a cFST, we start with the
initial state $q_S$ and repeatedly select a transition from $\delta(q,w)$,
where $q$ is the current state and $w$ the next input item. If there are
multiple such transitions (i.e., when $\sz{\delta(q,w)} > 1$), we try them
one by one via backtracking. As we move from state to state, we keep track of
the outputs in a buffer (column ``Produces'' in
Tab.~\ref{tab:item_expressions}). If we reach a final state after consuming
all input items, we add the buffered output to the set $G_{\cA}(T)$.

\textbf{Partial matches.}  The simulation algorithm only generates an output
when the entire input sequence is matched. If we are interested in matching
pattern expressions that occur somewhere in the input sequence instead, we
construct a cFST for $.^*E$ (instead of for $E$) and modify the above
simulation such that it adds the buffered output to $G_\cA(T)$ whenever a
final state is reached, whether or not the entire input has been
consumed.\footnote{This approach is more efficient than using expression
$.^*E.^*$ for constructing the cFST.}

\begin{figure}[t]
\usetikzlibrary{calc}
\centering
\tiny
\subfigure[FST]{
	\begin{tikzpicture}
		[->,>=stealth',
		shorten >=1pt,
		auto,
		node distance=2.8cm,
		semithick]
		\tikzstyle{every state}=[inner sep=0.8pt]
		\node[initial, initial text={}, state] (q0){$q_0$};
		\node[right=3mm of q0] (dummy) {};

		\node[state, above=3mm of dummy] (q1) {$q_1$}; 
		\node[state, below=3mm of dummy] (q2) {$q_2$};
		\node[state, right=6mm of q1] (q3) {$q_3$};
		\node[state, right=6mm of q2] (q4) {$q_4$};

		\node[state, right=21mm of q0] (q5) {$q_5$};
		\node[right=5mm of q5] (dummy2) {};
		\node[state, above=4mm of dummy2] (q6) {$q_6$}; 
		\node[state, below=4mm of dummy2] (q7) {$q_7$};
		\node[state, right=10mm of q6] (q8) {$q_8$};
		\node[state, right=10mm of q7] (q9) {$q_9$};

		\node[state, right=27mm of q5] (q10) {$q_{10}$};
		\node[accepting, state, right=6mm of q10] (q11) {$q_{11}$};

		\path 
			(q0) edge [bend left] node[above left]{$\epsilon$} (q1)
			(q0) edge [bend right] node[below left]{$\epsilon$} (q2)
			(q1) edge [above] node {$c$:$\epsilon$} (q3)
			(q2) edge [above] node {$d$:$\epsilon$} (q4)
			(q3) edge [bend left] node[near start]{$\epsilon$} (q5)
			(q4) edge [bend right] node[below,midway]{$\epsilon$} (q5)
			(q5) edge [bend left] node[above]{$\epsilon$} (q6)
			(q5) edge [bend right] node[below]{$\epsilon$} (q7)
			(q6) edge [bend left=70, above] node {$a_1$:$a_1$} (q8)
			(q6) edge [bend left=30, above] node {$a_1$:$A$} (q8)
			(q6) edge [above] node {$a_2$:$a_2$} (q8)
			(q6) edge [bend right=30, above] node {$a_2$:$A$} (q8)
			(q6) edge [bend right=60, above] node {$A$:$A$} (q8)
			(q7) edge [bend left=70, above] node {$b_{11}$:$B$} (q9)
			(q7) edge [bend left=30,above] node {$b_{12}$:$B$} (q9)
			(q7) edge [above] node {$b_{1}$:$B$} (q9)
			(q7) edge [bend right=30, above] node {$b_{2}$:$B$} (q9)
			(q7) edge [bend right=60,above] node {$B$:$B$} (q9)
			(q8) edge [bend left] node[near start,above=0mm]{$\epsilon$} (q10)
			(q9) edge [bend right] node[below]{$\epsilon$} (q10)
			(q10) edge [above] node {$e$:$\epsilon$} (q11)
			(q10) --node (bot) [minimum width=-2pt,inner sep=-1pt,above=1.6cm]{} (q5);
		;
		\draw[-] (q10) edge[out=90,in=0] (bot);
		\draw[->] (bot) edge[out=180,in=90] (q5);
		\node[above=0mm of bot] {$\epsilon$};
	\end{tikzpicture}	
	\label{fig:example_fst}
}
\subfigure[Compressed FST]{
\begin{tikzpicture}
[->,>=stealth',
shorten >=1pt,
auto,
node distance=2.8cm,
semithick]
\tikzstyle{every state}=[inner sep=0.8pt]  
\node[initial,initial text={},state] (q0){$q_0$};
\node[state, right=10mm of q0] (q1) {$q_1$};
\node[state, right=10mm of q1] (q2) {$q_2$};
\node[accepting, state, right=10mm of q2] (q3) {$q_3$};

\path
(q0) edge [bend left, above] node {$c$:$\epsilon$} (q1)
(q0) edge [bend right, below] node {$d$:$\epsilon$} (q1)
(q1) edge [bend left, above] node {$A$:$\$\hh A$} (q2)
(q1) edge [bend right, below] node {$B$:$B$} (q2)
(q2) edge [loop above, right] node {$A$:$\$\hh A$} ()
(q2) edge [loop below, above right] node {$B$:$B$} ()
(q2) edge [above] node {$e$:$\epsilon$} (q3)
;
\end{tikzpicture}
\label{fig:example_compressed_fst}
}
\caption{FST (a) and cFST (b) for $[c|d]([A^\gen\mid B_\force^\gen]^+)e$.}
\end{figure}

\textbf{Nondeterminism.}  Note that cFST simulation involves backtracking when
multiple transitions match the same input item and/or a transition has an
output label of form $\$\mhyphen w$ or $\$\mhyphen\top$. The standard way to
avoid non-determinism is to use some form of FST
determinization~\cite{Mohri:1997}. In general, these methods do not directly
apply to our FSTs because there are no ``sequential'' or even
``$p$-subsequential'' transducers for some pattern expressions (e.g.  $E =
[.^*(.)]^+$). In our implementation, we adapt the power construction
algorithm~\cite{Rabin:1959} and some other heuristics to reduce
non-determinism to the extent possible. The remaining non-determinism (if any)
did not lead to a bottleneck in our experimental study.


\section{Pattern Mining}\label{sec:pattern_mining} 

We now turn attention to mining $P$-frequent sequences from a sequence
database. We assume that subsequence predicate $P$ is described by a cFST
$\cA$ (e.g., obtained by translating a pattern expression).  We propose three
methods for mining $P$-frequent sequences: Na\"ive, \NAME-COUNT, and
\NAME-DFS. We assume that subsequence predicate $P$ is described by a cFST
$\cA$.

The na\"ive approach is to compute all $P$-generated sequences for each
input sequence, count how often each sequence has been obtained, and output
the ones that are frequent. \NAME-COUNT improves on the na\"ive approach by
only generating sequences that do not contain globally infrequent
items. Finally, \NAME-DFS is based on depth-first projection-based
methods~\cite{prefixspan,prefix_monotone} and is generally more efficient than
\NAME-COUNT when the set of $P$-generated sequences is large.

\subsection{Na\"ive Approach} \label{subsec:naive_approach}

The na\"ive ``generate-and-count'' approach is to compute $G_\cA(T)$ for each
input sequence $T\in \xD$ via cFST simulation and count how often each
sequence has been generated (cf.~Eq.~\eqref{eq:sup}). The na\"ive approach is
generally inefficient because it considers many globally infrequent sequences.
For example,  we obtain
\begin{align*}
G_{A_{ex}}(T_3)=&\{AAAB, AAa_2B, Aa_1AB, Aa_1a_2B, \label{ex1:T3} \\ &a_2AAB,
 a_2Aa_2B, a_2a_1AB, a_2a_1a_2B\} \nonumber
\end{align*}
for input sequence $T_3$, but only $AAAB$ and $Aa_1AB$ are $P$-frequent.

\subsection{\NAME-COUNT}\label{subsec:p-count}

\NAME-COUNT reduces the number of sequences that are generated and counted by
making use of item frequencies. In more detail, denote by
$f(w,\xD)=\sz{\{T\in\xD \mid w\sqsubseteq T\}}$ the \emph{frequency} of item
$w$. We say that item $w$ is \emph{frequent} if $f(w,\xD)\ge\sigma$. Similar
to many prior FSM algorithms, \NAME-COUNT first generates an \emph{f-list}
$F$, which contains all items along with their frequencies. For our example
database, we obtain f-list
\begin{IEEEeqnarray}{Rl}
F_{ex}=&\{A\cc6, e\cc6,B\cc6,a_1\cc6,d\cc3,b_2\cc3, b_1\cc2,c\cc2, \nonumber\\
&b_{12}\cc1,b_{11}\cc1,a_2\cc1\}. \label{eq:fex}
\end{IEEEeqnarray}

Note that the f-list is independent of the subsequence constraint and can be
precomputed. In \NAME-COUNT, we make use of the f-list to reduce the size of
$G_\cA(T)$.  Denote by
\[
 G_\cA^F(T)=\sset{S\in G_\cA(T) \mid \forall w\in S : f(w,\xD)\ge\sigma} 
\]
the subset of generated sequences that do not contain infrequent items.  For
$T_3$, we have $G_{\cA_{ex}}^{F_{ex}}(T_3) =\{\, AAAB, Aa_1AB\}$, which is
much smaller than the full set $G_{\cA_{ex}}(T_3)$ given above. \NAME-COUNT
proceeds as the na\"ive approach, but replaces $G_{\cA}(T)$ by $G_{\cA}^F(T)$
for each $T\in\xD$. Note that we do not fully compute $G_{\cA}(T)$ to obtain
$G_{\cA}^F(T)$, instead we directly compute the reduced set $G_\cA^F(T)$ by
adapting cFST simulation to work with the f-list. To do so, we stop exploring
a run as soon as an infrequent item is produced.

The correctness of \NAME-COUNT is established by
Lemma~\ref{lem:subsequence_program}, which states that FSTs specify
subsequence predicates, and the following observation.

\begin{lemma}
\label{lemma:item_monotonicity}
Let $P$ be a subsequence predicate and $S\in\Sigma^+$ be any sequence. Then
  for all $w\in S$, $f(w,\xD) \ge f_P(S,\xD)$.
\end{lemma}
\begin{IEEEproof} 
  Pick any $w\in S$ and input sequence $T\in\xD$ such that $S\in G_P(T)$.
  Since $P$ is a subsequence predicate, $S \sqsubseteq T$. Since $w\in S$, we
  have $w\sqsubseteq S$ and thus also $w\sqsubseteq T$. We obtain
\begin{IEEEeqnarray}{Rl}
    f_P(S,\xD) & = \sz{\sset{T\in\xD \mid S\in G_P(T)}} \IEEEnonumber\\
    & \le \sz{\sset{T\in\xD \mid w\sqsubseteq T}} = f(w,\xD) \IEEEnonumber 
\end{IEEEeqnarray}
\end{IEEEproof}

The lemma implies that $P$-frequent sequences must be composed of frequent
items. We thus can safely prune all sequences that contain infrequent items
from $G_{\cA}(T)$.

The pruning performed by \NAME-COUNT can substantially reduce the number of
candidate sequences. \NAME-COUNT is inefficient (and sometimes infeasible),
however, if pruning is not sufficiently effective and the sets $G_\cA^F(T)$
are very large. The \NAME-DFS algorithm, which we present next, addresses such
cases.

\subsection{\NAME-DFS}\label{subsec:p-incr}

DESQ-DFS adapts the pattern-growth framework of PrefixSpan~\cite{prefixspan}
to FSTs. Pattern growth approaches arrange the output sequences in a tree, in
which each node corresponds to a sequence $S$ and is associated with a
\emph{projected database}, which consists of the set of input sequences in
which $S$ occurs. Starting with an empty sequence and the full sequence
database, the tree is built recursively by a performing series of
\emph{expansions}. In each expansion, a frequent sequence $S$ (of $l$ items)
is expanded to generate sequences (of $l+1$ items) with prefix $S$, their
projected databases, and their supports. In what follows, we describe how we
adapt these concepts to mine $P$-frequent sequences; the corresponding
algorithm for cFSTs is shown as Alg.~\ref{alg:p-incr} and illustrated on our
running example in Fig.~\ref{fig:desq-incr}.

\textbf{Projected databases.} For a sequence $S$, we store in its
projected database the state of the simulations of $\cA$ on all input sequences
that generate $S$ as a partial output. We refer to such a state as a
\emph{snapshot} for $S$. The snapshot concisely describes which items have been
consumed, which state the FST simulation is in, and which output has been
produced so far. In more detail, suppose that we simulate an $\cA$ on input
sequence $T=t_1\dots t_n$. Consider a partial run $p=p_1\dots p_{m}$ obtained
after $m\le n$ steps. We generated output $S=O(p)$ and, under our running
assumption that $\cA$ does not contain $\eps$-transitions, consumed prefix
$T'=t_1\dots t_{m}$ of $T$ at this time. If the output item of the last
transition $p_m$ is not empty (and thus agrees with the last item of $S$), we
say that triple $T[pos@q]$ is a \emph{snapshot} for $S$, where $pos=m+1$ is
the position of next input item and $q$ is the last state in $p$ (current
state of $\cA$). The \emph{projected database} for $S$ consists of all
snapshots for $S$ and is given by
\begin{align*}
\Proj_{\cA}(S,\xD)=\{\,T[pos@q] \mid\,&T\in\xD, T[pos@q]\text{ is a snapshot} \\
  &\text{for $S$ on $\cA$}\,\}.
\end{align*}
Fig.~\ref{subfig:projected_databases} shows
some projected databases associated with some sequences for our running
example. For example, we obtained the partial output $a_1$ only from input
sequences $T_1$, $T_4$, and $T_6$. In each case, we consumed two items
(next item is at position 3) and ended in state $q_2$.
We refer to the number of input sequences that can
generate $S$ as a partial output as the \emph{prefix support} of $S$:
\[
\Presup_\cA(S,\xD) = \{T \mid \exists pos,q:T[pos@q] \in \Proj_\cA(S,\xD)\}.
\]
In our example, $\Presup_{\cA_{F{\ref{fig:desq-incr}}}}(a_1,\xD_{ex}) =
\sset{T_1,T_4,T_6}$. Note that even if an input sequence has multiple snapshots
for $S$, it contributes only once to the prefix support.


\begin{algorithm}[tb]
\algsetup{linenosize=\footnotesize}
\footnotesize
\begin{algorithmic}[1]
  \REQUIRE $\xD$, cFST $\cA = (Q,q_S,Q_F,\Sigma,\Delta)$, $\sigma$, f-list $F$
  \ENSURE $P$-frequent sequences for $\cA$ in $\xD$
  \STATE $S\gets \epsilon$ \COMMENT{create root node; initially fields $S\ldot\List=S\ldot\PSup=\emptyset$} \label{p_incr:init}
  \STATE $S\ldot\List \gets \sset{T_1[1@q_S],\ldots,T_{\sz{\xD}}[1@q_S]}$ \label{p_incr:init_1}
  \STATE \textsc{Expand}$(S)$ \label{p_incr:expand_1}
  \STATE
  \STATE \textbf{void} \textsc{Expand}$(S)$:
  \FORALL[simulate cFST for all snapshots]{$T[pos@q] \in S\ldot\List$} \label{p_incr:scan_list_begin}
    \STATE \textsc{IncStep}$(T,pos,q,S$) \label{p_incr:scan_list_end}
  \ENDFOR
  \SIF{$\sz{S\ldot\PSup} \ge \sigma$} \STHEN{\textbf{yield} ($S$,$\sz{S\ldot\PSup}$)} \COMMENT{Output if $P$-frequent} 
    \label{p_incr:output}
  \FORALL[expand if prefix support large enough]{$S'\in S\ldot\Children$} \label{p_incr:for_all_chilren}
    \SIF{$\sz{\sset{T\mid \exists pos,q\colon T[pos@q]\in S\ldot\Proj}} \ge \sigma$} \STHEN{\textsc{Expand}$(S')$}\label{p_incr:expand_2}
  \ENDFOR
  \STATE 
  \STATE \textbf{void} \textsc{IncStep}$(T,pos,q,S$):		\label{p_incr:lockstep_begin} \COMMENT{simulate until an item is produced}
  \IF{$q \in Q_F$ \AND $pos > \sz{T}$ \AND $S\neq\eps$}\label{p_incr:sup_p_if} 
   \STATE $S\ldot\PSup \gets S\ldot\PSup \cup \sset{T}$  \COMMENT{initially empty} \label{p_incr:sup_p}
  \ENDIF
    \FORALL{$(out,q_{to}) \in \delta(q, t_{pos}$)} 
      \SWITCH{out}
      \CASE{$\epsilon$}
        \STATE \textsc{IncStep}$(T,pos\pp1,q_{to},S$) \label{p_inc:case_e}
      \CASE{$w$} 
        \SIF{$f(w,\xD) \ge \sigma$} \STHEN{\textsc{Append}($S$, $w$, $T$, $pos\pp1$, $q_{to}$)}\label{p_incr:case_w}
      \CASE{$\$$} 
        \SIF{$f(t_{pos},\xD) \ge \sigma$} \STHEN{\textsc{Append}($S$, $t_{pos}$, $T$, $pos\pp1$, $q_{to}$)} \label{p_incr:case_dollar}
      \CASE{$\$\mhyphen x, x \in \Sigma \cup \sset{\top}$} 
        \FORALL{$w' \in \anc(t_{pos}) \cap \desc(x)$}\label{p_incr:case_dollar_x_begin}
          \SIF{$f(w',\xD) \ge \sigma$} \STHEN{\textsc{Append}($S$, $w'$, $T$, $pos\pp1$, $q_{to}$)} \label{p_incr:case_dollar_x}
        \ENDFOR
      \ENDSWITCH
    \ENDFOR 	\label{p_incr:lockstep_end}
    \STATE
    \STATE \textbf{void} \textsc{Append}$(S, w, T, pos, q)$: \label{p_incr:append_begin}
    \STATE $S\ldot\Children \gets S\ldot\Children \cup \sset{Sw}$ \COMMENT{node $Sw$ is created if new}
    \STATE $Sw\ldot\List \gets Sw\ldot\List \cup \sset{T[pos@q]}$ \COMMENT{initially empty} \label{p_incr:append_end}
  \end{algorithmic}
  \caption{\NAME-DFS}
  \label{alg:p-incr}
\end{algorithm}


\textbf{Expansions.}  We start with root node $\eps$ and all
snapshots for $\eps$ (lines~\ref{p_incr:init} and~\ref{p_incr:init_1}) and
then perform a series of expansions (lines~\ref{p_incr:expand_1} and
\ref{p_incr:expand_2}). In each expansion, we scan the projected database
sequentially. For each snapshot $T[pos@q]$
(lines~\ref{p_incr:scan_list_begin}--\ref{p_incr:scan_list_end}), we resume
the FST for $T$ at item $t_{pos}$ in state $q$ (via \textsc{IncStep}, lines
\ref{p_incr:lockstep_begin}--\ref{p_incr:lockstep_end}). The transducer is
stopped as soon as an output item is produced or the entire input is consumed.
In the former case, suppose we produce item $w$ after consuming $k$ more input
items from $T$ and thereby reach state $q'$. We then add the new snapshot
$T[pos\pp k@q']$ to the projected database of child node $Sw$
(lines~\ref{p_incr:case_w},~\ref{p_incr:case_dollar}, and
\ref{p_incr:case_dollar_x}). In the later case, if we end up in a final state
(lines~\ref{p_incr:sup_p_if}--\ref{p_incr:sup_p}), we conclude that
$T\in\Sup_\cA(S,\xD)$ (see below). For example, both snapshots of $a_1B$ reach
final state $q_3$ by consuming all input items and without producing further
output, so that $a_1B\ldot\Sup=\sset{T_1,T_4}$.

\textbf{Pruning.} The above expansion procedure allows us to prune
partial sequences as soon as it becomes clear that they cannot be expanded to
a $P$-frequent sequence. We use two pruning techniques. First, as in \NAME-
COUNT, we consider item $w$ only if it is frequent; otherwise, we ignore the
new snapshot. For example, when expanding $a_1$, we do not create nodes for
sequences that contain infrequent items; e.g., $a_1b_{12}$ has snapshot
$T_1[4@q_2]$ but contains infrequent item $b_{12}$ (see Eq.~\eqref{eq:fex}).
Second, we expand only those nodes $S$ that have a sufficiently large prefix
support---i.e., $\Presup_\cA(S,\xD)\ge\sigma$---and stop as soon as there is
no such node anymore.  For example, we do not expand node $a_1a_1$ because it
contains only one snapshot, but we require snapshots from at least $\sigma=2$
different input sequences.


\begin{figure}[t]
\addtolength{\abovecaptionskip}{-6pt}
	\centering
	\tikzstyle{expansion}=[inner sep=1pt,draw,fill=black]
	\tikzstyle{empty}=[fill=black]
	\tikzstyle{fprefix}=[inner sep=1pt,draw,rounded corners,minimum width=4mm,minimum height=4mm]
	\tikzstyle{iprefix}=[inner sep=1pt,draw,rounded corners, dashed, minimum width=4mm,minimum height=4mm]
	\tikzstyle{loutput}=[double]
	\tikzstyle{lexpansion}=[inner sep=2pt,draw,fill=black]
	\tikzstyle{empty}=[fill=black]
	\tikzstyle{lfprefix}=[inner sep=2pt,draw,rounded corners=1mm,minimum width=3mm,minimum height=3mm]
	\tikzstyle{liprefix}=[inner sep=2pt,draw,rounded corners=1mm, dashed, minimum width=3mm,minimum height=3mm]
	\tikzstyle{output}=[double]
	\subfigure[Search space]{
	\label{subfig:search_space}
	\tiny
	\begin{tikzpicture}
		\node[fprefix] (ep) {$\epsilon$};
		\node[expansion, below=2mm of ep] (e1) {};
		\draw[-] (ep) edge node[left]{E1} (e1);

		\node[below=4mm of e1] (dum1) {};
		\node[fprefix, left=15mm of dum1] (a1) {$a_1$};
		\node[fprefix, right=15mm of dum1] (A) {$A$};
		\draw[-] (e1) edge (a1.north east) edge (A.north west);

		\node[expansion, below=2mm of a1] (e2) {};
		\node[output,fprefix,below=2mm of e2] (a1B) {$a_1B$};
		\node[iprefix, left=6mm of a1B] (a1A) {$a_1A$};
		\node[iprefix, right=6mm of a1B] (a1a1) {$a_1a_1$};
		\draw[-] (a1) edge node[left]{E2} (e2);
		\draw[-] (e2) edge (a1A.north east) edge (a1B.north) edge (a1a1.north west);

		\node[expansion, below=2mm of a1B] (e3) {};
		\draw[-] (a1B) edge node[left]{E3} (e3);

		\node[expansion, below=2mm of A] (e4) {};
		\node[output,fprefix,below=2mm of e4] (AB) {$AB$};
		\node[fprefix, left=6mm of AB] (AA) {$AA$};
		\node[fprefix, right=6mm of AB] (Aa1) {$Aa_1$};
		\draw[-] (A) edge node[left]{E4} (e4);
		\draw[-] (e4) edge (AA.north east) edge (AB.north) edge (Aa1.north west);

		\node[expansion, below=2mm of AA] (e5) {};
		\draw[-] (AA) edge node[left]{E5} (e5);

		\node[fprefix,below left=2mm of e5] (AAA) {$AAA$};
		\node[iprefix,below right=2mm of e5] (AAa1) {$AAa_1$};
		\draw[-] (e5) edge (AAA.north east) edge (AAa1.north west);

		\node[expansion, below=2mm of AAA] (e6) {};
		\draw[-] (AAA) edge node[left]{E6} (e6);
		\node[fprefix,output,below=2mm of e6] (AAAB) {$AAAB$};
		\draw[-] (e6) edge (AAAB);
		\node[expansion, below=2mm of AAAB] (e7) {};
		\draw[-] (AAAB) edge node[left]{E7} (e7);

		\node[expansion, below=2mm of AB] (e8) {};
		\draw[-] (AB) edge node[left]{E8} (e8);

		\node[expansion, below=2mm of Aa1] (e9) {};
		\draw[-] (Aa1) edge node[left]{E9} (e9);

		\node[fprefix,below left=2mm of e9] (Aa1A) {$Aa_1A$};
		\node[iprefix,below right=2mm of e9] (Aa1a1) {$Aa_1a_1$};
		\draw[-] (e9) edge (Aa1A.north east) edge (Aa1a1.north west);

		\node[expansion, below=2mm of Aa1A] (e10) {};
		\draw[-] (Aa1A) edge node[left]{E10} (e10);
		\node[fprefix,output,below=2mm of e10] (Aa1AB) {$Aa_1AB$};
		\draw[-] (e10) edge (Aa1AB);
		\node[expansion, below=2mm of Aa1AB] (e11) {};
		\draw[-] (Aa1AB) edge node[left]{E11} (e11);

		\node[draw,font=\scriptsize, minimum width=3.7cm, text depth=1.55cm] at(-2.5,-3.5) (legend){};
		\node[lfprefix,loutput] at ([xshift=2.5mm,yshift=-1.5em]legend.north west) (pf){};
		\node[lfprefix, below=1mm of pf] (f) {};
		\node[liprefix, below=1mm of f](if) {};
		\node[lexpansion, below=1mm of if] (ex) {};
		\node[font=\scriptsize,right=2mm of ex] {Expansions};
		\node[font=\scriptsize,right=1mm of pf] {$P$-frequent, expanded};
		\node[font=\scriptsize,right=1mm of f] {Not $P$-frequent, expanded};
		\node[font=\scriptsize,right=1mm of if] {Not $P$-frequent, not expanded};

	\end{tikzpicture}
	}
	\subfigure[Some projected databases, prefix supports, and supports]{
	\label{subfig:projected_databases}
	\footnotesize
		\newcolumntype{R}{@{\extracolsep{6pt}}c@{\extracolsep{1pt}}}%
		\renewcommand{\tabcolsep}{1pt}
		\begin{tabular}{llRR}
			\toprule
			$S$ & $S\ldot\Proj$ & $\sz{S\ldot\Presup}$ & $\sz{S.\PSup}$\\
			\midrule
			$\epsilon$ & $\langle T_1[1@q_0],T_2[1@q_0],T_3[1@q_0],T_4[1@q_0],$  & 6 & 0\\
			& \hfill $T_5[1@q_0],T_6[1@q_0] \rangle$ & & \\
			$a_1$ & $\langle T_1[3@q_2],T_4[3@q_2],T_6[3@q_2]\rangle$ & 3 & 0\\
			$a_1A$ & $\langle T_6[4@q_2] \rangle$ & 1 & 0\\
			$a_1B$ & $\langle T_1[3@q_2], T_4[3@q_2] \rangle$ & 2 & 2\\
			$a_1a_1$ & $\langle T_6[4@q_2] \rangle$ & 1 & 0\\
			\bottomrule
		\end{tabular}
	}
\caption{Illustration of \NAME-DFS for $\xD_{ex}$, $\cA_{F{ex}}$, and $\sigma=2$}
\label{fig:desq-incr}
\end{figure}

\textbf{Correctness.} Note that the size of the prefix support is monotonically
decreasing as we go down the tree but always stays at least as large as the
support. This property, which we establish next, is key to the correctness of
\NAME-DFS.

\begin{lemma}
\label{lem:prefix_monotonicity} 
For any sequence $S \in \Sigma^*$ and item $w \in \Sigma$, we have
  $ \Presup_\cA(Sw,\xD)\subseteq \Presup_\cA(S,\xD)$.
\end{lemma}
\begin{IEEEproof}
Pick any $S \in \Sigma^*$, $w\in \Sigma$, and $T=t_1\dots t_n\in\xD$ with
$T \in \Presup_\cA(Sw,\xD)$. Then there exists a run $p=p_1\dots p_m$ for
prefix $T'=t_1\dots t_m$ and some $m\le n$ such that $O(p)=Sw$. Recall that
inputs (outputs) are consumed (generated) from left to right. We
conclude that there exists some $m'<m$ such that run $p'=p_1\dots p_{m'}$
satisfies $O(p')=S$. Pick the shortest such run; then $p_{m'}$ outputs the
last item of $S$. Since $p'$ is additionally a run for $t_1\dots t_{m'}$,
which is a prefix of $T$, we conclude that $T\in \Presup_\cA(S,\xD)$.
\end{IEEEproof}

We now establish the correctness of \NAME-DFS.

\begin{theorem}
  \NAME-DFS outputs each $P$-frequent sequence $S\in\Sigma^+$ with frequency
  $f_P(S,\xD)$. No other sequences are output.
\end{theorem}
\begin{IEEEproof}
  Let $\cA = (Q, q_S, Q_F, \Sigma, \Delta)$ be an FST and pick any sequence
  $S\in\Sigma^+$. We start with showing that Alg.~\ref{alg:p-incr} correctly
  computes the $P$-support of $S$ when expanding node $S$, i.e.,
  $S\ldot\Sup=\PSup_\cA(S,\xD)$ after the expansion. First pick any $T\in
  \PSup(S,\xD)$ with $T=t_1\dots t_n$. Then there is an accepting run
  $p=p_1\dots p_n$ for $T$. By arguments as in the proof of
  Lemma~\ref{lem:prefix_monotonicity}, there must be a smallest run
  $p'=p_1\dots p_m$, $m\le n$, such that $O(p')=S$ as well. Let $q_m$ ($q_n$) be
  the state reached in transition $p_m$ ($p_n$). We conclude that snapshot
  $T[pos@q_m]\in\Proj_\cA(S,\xD)$, where $pos=m+1$, and thus
  $T\in\Presup(S,\xD)$. Since by definition $p_{m+1}\dots p_n$ must output
  $\eps$, Alg.~\ref{alg:p-incr} follows transitions $p_{m+1}\dots p_n$ without
  stopping when resuming snapshot $T[pos@q_m]$. By doing so, it consumes all the
  remaining items $t_{m+1}\dots t_n$ of $T$ and reaches final state $q_n$. It
  thus includes $T$ into $S\ldot\PSup$
  (lines~\ref{p_incr:sup_p_if}--\ref{p_incr:sup_p}).  Now pick $T\notin
  \PSup_\cA(S,\xD)$. Since there is no accepting run for $T$,
  Alg.~\ref{alg:p-incr} cannot reach a final state after consuming $T$ so that
  it does not include $T$ into $S\ldot\PSup$. Putting both together,
  $S\ldot\Sup=\PSup_\cA(S,\xD)$ after expanding $S$, as desired. We conclude
  that Alg.~\ref{alg:p-incr} computes the correct frequency
  $f_P(S,\xD)=\sz{\PSup_\cA(S,\xD)}$. $S$ is output only if it is $P$-frequent
  (line~\ref{p_incr:output}). Note that for $S=\eps$, we have
  $\eps\ldot\Sup=\emptyset$ (see line \ref{p_incr:sup_p_if}) so that $\eps$ is
  not output.

  Let $S\in\Sigma^+$ be a $P$-frequent sequence. It remains to show that
  Alg.~\ref{alg:p-incr} reaches and expands node $S$. First observe that for any
  prefix $S'$ of $S$, we have
  \[
  \Presup(S',\xD)\supseteq\Presup(S,\xD)\supseteq\Sup(S,\xD).
  \]
  Here the first inclusion follows from Lemma~\ref{lem:prefix_monotonicity}, and
  the second inclusion follows from the above arguments. Since $S$ is
  $P$-frequent, we have $\sz{\Sup(S,\xD)}\ge \sigma$, which implies
  $\sz{\Presup(S',\xD)}\ge \sigma$. Since every node on the path from $\eps$ to
  $S$ corresponds to a prefix of $S$, Alg.~\ref{alg:p-incr} does not prune any
  of these nodes due to too low prefix support (line~\ref{p_incr:expand_2}). To
  complete the proof, recall that $S$ cannot contain an infrequent item by
  Lemma~\ref{lemma:item_monotonicity}. Thus none of the nodes on the path from
  $\eps$ to $S$ are pruned due to too low item frequency either (lines
  \ref{p_incr:case_w}, \ref{p_incr:case_dollar}, or
  \ref{p_incr:case_dollar_x}). We conclude that Alg.~\ref{alg:p-incr} reaches
  and expands node $S$.
\end{IEEEproof}

To improve efficiency, our actual implementation of Alg.~\ref{alg:p-incr} does
not explicitly compute supports and prefix supports but directly counts their
sizes.


\section{Experimental Evaluation} \label{sec:experiments}

We conducted an experimental study on three publicly available real-world
datasets: a collection of text documents (for text mining), a collection of
product reviews (for customer behavior mining), and a collection of protein
sequences. Our goal was to investigate whether pattern expressions are
sufficiently powerful to express prior and new subsequence constraints, whether
\NAME's algorithms are efficient, and how they perform relative to each other
and to prior algorithms.

\textbf{Summary of our results.} (1) Many subsequence constraints can
be expressed with pattern expressions. (2) \NAME{} has acceptable overhead
over state-of-the-art specialized sequence miners for common subsequence
constraints. (3) \NAME-COUNT was consistently faster than Na\"ive. (4) \NAME-
COUNT and \NAME-DFS had similar performance in cases where the average number
of $P$-subsequences per input sequence was small. (5) When many
subsequences per input are generated, \NAME-DFS was more than an order of
magnitude faster than \NAME-COUNT and Na\"ive. (6) cFSTs sped up pattern
matching by multiple orders of magnitude when compared to the state-of-the-art
FST library OpenFST. Our results indicate that \NAME{} is a suitable general-purpose
system for a wide range of subsequence constraints.

\subsection{Experimental Setup}

\begin{table}[t] 
\newcolumntype{X}{@{\hspace{.2cm}}r}%
	\caption{Dataset statistics}
	\centering
	\begin{tabular}{l@{\hspace{.2cm}}rrXX}
		\toprule
		& & NYT & AMZN & PRT\\
		\midrule
		Sequence& \# Sequences & $\numprint{21590967}$ & $\numprint{6643666}$ & $\numprint{103120}$\\
		database& Avg. length & $\numprint{19.9}$ & $\numprint{4.5}$ & $\numprint{482}$ \\
		& Max. length & $\numprint{5042}$ & $\numprint{25630}$ & $\numprint{600}$ \\
		& Total items & $\numprint{430279662}$ & $\numprint{29667966}$ & $\numprint{49729890}$ \\
		& Distinct items & $\numprint{3975859}$ & $\numprint{2374096}$ & $\numprint{25}$\\ 
		\midrule
		Hierarchy
		& Total items & $\numprint{4136774}$ & $\numprint{2385775}$ & $\numprint{103120}$\\ 
		& Leaf items & $\numprint{3901118}$ & $\numprint{2371522}$ & $\numprint{103120}$ \\ 
		& Interm. items & $\numprint{235633}$ & $\numprint{11630}$ & 0\\ 
		& Root items & $\numprint{23}$ & $\numprint{2623}$ & $\numprint{103120}$\\ 
		& Max. depth & $\numprint{3}$ & $\numprint{8}$ & 1\\ 
		& Avg. fan-out & $\numprint{17.5}$ & $\numprint{204}$ & 0\\ 
		& Max. fan-out & $\numprint{1505913}$ & $\numprint{332723}$ & 0\\ 
		\bottomrule
	\end{tabular}
	\label{tab:dataset_characterstics}
\end{table}


\begin{table*}[t]
\newcolumntype{R}{@{\extracolsep{-95pt}}>{\it}r@{\extracolsep{4pt}}}%
\caption{
Example pattern expressions for traditional sequence mining
(${T_1}$--${T_3}$), protein sequence mining ($P_1$--$P_4$), IE and NLP
applications (${N_1}$--${N_5}$) and customer behavior mining applications
(${A_1}$--${A_4}$)}
\centering
\begin{tabular}{lRr}
\toprule
Pattern expression & \multicolumn{1}{r}{Description} & Example patterns from NYT dataset (frequency) \\
\midrule
${T_1}$: (.)\{1,$\lambda$\} & $n$-grams of up to $\lambda$ words & green tea (337), editor in chief (3275) \\
${T_2}$: (.)[.\{0,$\gamma$\}(.)]\{1,$\lambda-1$\} & Skip $n$-grams with gap at most $\gamma$ words and of up to length $\lambda$ & flight from to (758), son of and of (15896)
\\
${T_3}$: ($.^\gen$)\{1,$\lambda$\} &  Generalized $n$-grams of up to $\lambda$ words & NOUN\,PREP\,DET\,NOUN (4.2M), PERSON be NOUN (2199)\\ 
\midrule
& & Example patterns from PRT dataset (frequency) \\
\midrule
$P_1$: ([S$\mid$T])\,.$^*$(.)\,.$^*$([R$\mid$K]) & subsequences that match RE$\equiv$[S$\mid$T]\,.\,[R$\mid$K] & S\,L\,R (\numprint{103093}), T\,A\,K (\numprint{102941}), S\,A\,K (\numprint{102946})\\
$P_2$: ([I$\mid$V])\,.$^*$(D)\,.$^*$(L)\,.$^*$(G)\,.$^*$(T)\,.$^*$([S$\mid$T])\,.$^*$(.)\,.$^*$([S$\mid$C]) & subsequences that match  &  I\,D\,L\,G\,T\,T\,L\,S (\numprint{102975}), V\,D\,L\,G\,T\,S\,T\,C (\numprint{92662}) \\
                                        & RE$\equiv$[I$\mid$V]\,D\,L\,G\,T\,[S$\mid$T]\,.[S$\mid$C] & V\,D\,L\,G\,T\,S\,D\,S (\numprint{102901})  \\
$P_3$: ([S$\mid$T]\,.\,[R$\mid$K]) & contiguous subsequences that match RE$\equiv$[S$\mid$T]\,.\,[R$\mid$K] & S\,L\,R (\numprint{14995}), T\,A\,K (\numprint{8840}), S\,A\,K (\numprint{10397})\\
$P_4$: ([S$\mid$T]\,.\,.\,[D$\mid$E]) & contiguous subsequences that match RE$\equiv$[S$\mid$T]\,.\,.\,[D$\mid$E] & S\,D\,L\,E (\numprint{2015}), T\,L\,E\,E (\numprint{2329}), S\,G\,L\,D (\numprint{1054}) \\ 
\midrule
& & Example patterns from NYT dataset (frequency) \\
\midrule
${N_1}$: ENTITY\,(VERB$^+$\,NOUN$^+$?\,PREP?)\,ENTITY & Relational phrase between entities & lives in (847), is being advised by (15), has coached (10)\\
${N_2}$: {(ENTITY$^\gen$\,VERB$^+$\,NOUN$^+$?\,PREP?\,ENTITY$^\gen$)} & Typed relational phrases &  ORG headed by ENTITY (275), PER born in LOC (481)\\
$N_3$: (ENTITY$^\gen$\,be$_\force^\gen$)\,DET?\,(ADV?\,ADJ?\,NOUN) & Copular relation for an entity & PER be novelist (165), LOC be great place (38),\\ 
${N_4}$: $(.^\gen)\{3\}$\,NOUN & Generalized 3-grams before a noun &  NOUN PREP DET (\numprint{4223219}), DET ADV ADJ (\numprint{350005}) \\ 
${N_5}$: ([$.^\gen \,. \,\,.]|[. \,\, .^\gen \,.]|[.\,\, . \,\,.^\gen]$) & Generalized 3-grams, where at most one item is generalized & the ADJ human (\numprint{1238}), for DET book (\numprint{1704})\\
\midrule
& & Example patterns from AMZN dataset (frequency) \\
\midrule
${A_1}$: (Electr$^\gen$)[.\{0,2\}(Electr$^\gen$)]\{1,4\} & Generalized sequences of (up to 5) electronic items, & ``Mice'', ``Keyboards'', ``Computers \& Accessories'' (556),\\
& which are at most 2 items apart in the input sequences & ``MP3 Players'', ``Headphones'' (814) \\
${A_2}$: (Book)[.\{0,2\}(Book)]\{1,4\}& Sequences of books &  ``The Bourne Supremacy'', ``The Bourne Ultimatum'' (16)\\
${A_3}$: DigitalCamera[.\{0,3\}($.^\gen$)]\{1,4\} & Type of products bought after a digital camera &  ``Lenses'', ``Tripods'' (11), ``Batteries'', ``SD Cards'' (12)\\
${A_4}$: (MInstr$^\gen$)[.\{0,2\}(MInstr$^\gen$)]\{1,4\} & Generalized sequences of musical instruments& ``Saxophones'', ``Bags \& Cases'', ``Instr. Accessories'' (127) \\
\bottomrule
\end{tabular}
\label{xtab:exp_pattern_expressions}
\end{table*}



\textbf{Datasets.} Tab.~\ref{tab:dataset_characterstics} summarizes our
datasets. NYT is a subset of the New York Times corpus~\cite{NYTcorpus} and
contains news articles. We generated an item hierarchy using annotations from
the Stanford CoreNLP tools. The NYT hierarchy consists of named entities,
which generalize to their type (PERSON, ORGANIZATION, LOCATION, MISC) and then
to ENTITY, and of words, which generalize to their lemma and then to their
part-of-speech tag.  For example,
``Maradona''$\Rightarrow$PERSON\allowbreak$\Rightarrow$\allowbreak{}ENTITY and
``is''$\Rightarrow$``be''\allowbreak{}$\Rightarrow$VERB.

AMZN is a dataset of Amazon product reviews~\cite{Amazon} from which we
extracted sequences of products (ordered by review timestamps) for each user.
We used the Amazon product hierarchy as our item hierarchy. For example,
``Canon 5D'' $\Rightarrow$``Digital Cameras''$\Rightarrow$``Camera \&
Photo''$\Rightarrow$ ``Electronics''.

PRT is a dataset of protein sequences~\cite{sma} composed of 25 amino acid
codes (items). The hierarchy is flat, i.e., there are no generalizations.

\textbf{Implementation and setup.} We implemented \NAME{} in Java
(JDK 1.8;
\url{http://dws.informatik.uni-mannheim.de/en/resources/software/desq/}). We
used ANTLR4 to generate a parser for pattern expressions. The cFST is
constructed from the resulting parse tree. To measure the overhead of DESQ for
common subsequence constraints, we compared it against state-of-the- art
methods. For length and gap constraints, we used (1) C++ implementation of
cSPADE~\cite{cspade} from the author, (2) our implementation of SPADE in Java
that additionally handles hierarchy constraints, (3) our implementation of
prefix- growth~\cite{prefix_monotone} in Java.  For RE constraints, we used
(1) prefix-growth and a C++ executable of SMA~\cite{petri_net} obtained from
the authors. To evaluate cFSTs we compared it against state-of- the-art FST
library OpenFST 1.5.0~\cite{openFst-download}.

We preprocessed the datasets to compute the f-list and assign integer
identifiers to each item. Item identifiers were assigned in descending order
of item frequency, thus a more frequent item received a smaller item
identifier. In our implementations, we encoded the sequence database compactly
as arrays of item identifiers and use variable-length byte encoding to
compress projected databases. Experiments on the NYT and AMZN datasets were
performed on a machine with two Intel(R) Xeon(R) CPU E5-2640 v2 processors and
128GB of RAM running CentOS Linux 7.1. Experiments on the PRT dataset were
performed on a machine equipped with Intel Core i7-4712HQ and 16GB RAM running
Windows 10.  We used a different setup for the PRT dataset as the SMA
implementation is provided as a Windows binary only. All experiments were run
single-threaded.

\textbf{Methodology.} For each experiment, we report the performance in
terms of the total wall-clock time between launching the mining task and
receiving the final result (including I/O). All measurements were averaged
over three independent runs. Unless stated otherwise, all methods produced the
same result.

\begin{figure*}[t]
\centering
\subfigure[\scriptsize{NYT (length, gap, hierarchy)-constraints}]{
	\label{fig:traditional-constraints-time}
	\includegraphics[scale=0.8]{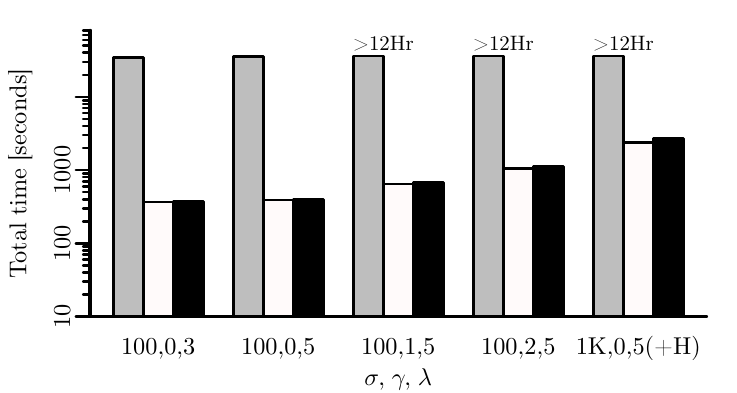}
}
\subfigure[\scriptsize{NYT (length, gap, hierarchy)-constraints}]{
	\label{fig:traditional-constraints-memory}
	\includegraphics[scale=0.8]{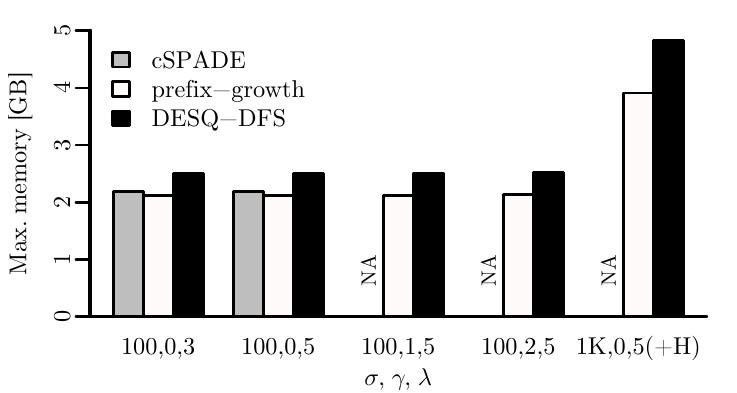}
}
\subfigure[\scriptsize{RE-constraints}] {
\label{fig:re-constraints}
\includegraphics[scale=0.8]{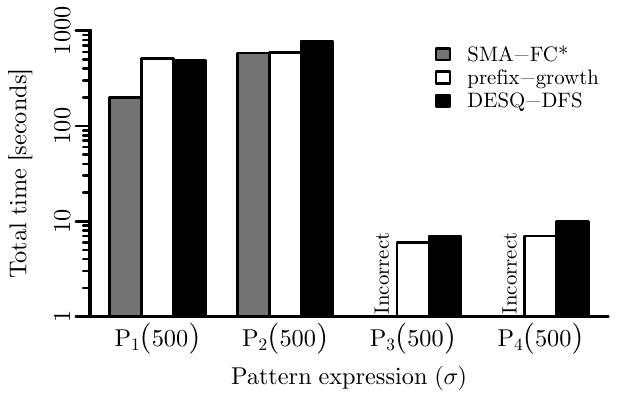}
}
\caption{Overhead of DESQ for common subsequence constraints.}
\end{figure*}

\subsection{Overhead of DESQ for Common Subsequence Constraints}

We first investigated the overhead of DESQ compared to specialized miners for
prior subsequence constraints. In particular we considered traditional as well
as RE constraints.

\textbf{Traditional constraints.} We considered length and gap
constraints as well as item hierarchies. We map these constraints to pattern
expressions and obtain $T_1$--$T_3$ of
Tab.~\ref{xtab:exp_pattern_expressions}. The expressions are parameterized by
maximum-length parameter $\lambda$ and/or maximum-gap parameter $\gamma$. We
used the NYT dataset and ran FSM for different configurations of increasing
difficulty w.r.t. output size. The results are shown in
Fig.~\ref{fig:traditional-constraints-time} using log-scale. For $n$-grams
(first two groups), we observed that DESQ-DFS was up to two orders of
magnitude faster than cSPADE. We only show the result for our own cSPADE
implementation; the original C++ implementation was significantly slower. For
example, for mining 10\% of NYT, the original cSPADE implementation took more
than 3 hours whereas our implementation took 400 seconds. Both cSPADE
implementations were significantly slower than prefix-growth and DESQ-DFS,
however, because cSPADE follows a candidate-generation-and-test approach and
suffers from an excessive number of generated candidates. To keep our study
manageable, we stopped cSAPDE after 12 hours. Compared to prefix-growth,
DESQ-DFS had negligible overhead (less than 2.5\%). For gap constraints (third
and fourth group), DESQ-DFS was competitive and had an overhead of less than
10\% over prefix- growth.  This overhead is expected as pattern expressions
for gap constrains have uncaptured wildcards (cf.~$T_2$ in
Tab.~\ref{xtab:exp_pattern_expressions}), which increases nondeterminism in
the corresponding cFSTs and thus leads to more snapshots. For generalized
$n$-grams (last group), where we additionally considered item hierarchies, the
overhead was slightly more pronounced (up to 13\%). Here the amount of
backtracking performed by DESQ increased with the depth of hierarchy (cf.
line~\ref{p_incr:case_dollar_x_begin} of Alg.~\ref{alg:p-incr} and discussion
in Sec.~\ref{sec:pattern_matching}).

We also investigated the overhead in terms of memory consumption. The results
are shown in Fig.~\ref{fig:traditional-constraints-memory}. For cSPADE, we
report the maximum size of the inverted index and for prefix-growth and
DESQ-DFS, we report the maximum size of the projected database. For $n$-grams
and gap-constraints, DESQ-DFS had an overhead of up 18\% and for generalized
$n$-grams up to 23\%. The overhead is unavoidable as for DESQ-DFS, we need to
store cFST snapshots compared to only positional information as in
prefix-growth and cSPADE. We may, however, improve memory consumption by
swapping projected databases to disk~\cite{prefixspan}. We also ran the above
experiments for the AMZN dataset (not shown here) and observed a similar
behavior.

\textbf{RE constraints.} In this set of experiments, we evaluated
the efficiency of \NAME{} for mining frequent subsequences (all or contiguous)
that match a RE. Our pattern expressions allows us to express REs with their
equivalent pattern expressions (cf. Tab~\ref{tab:existing_relevance} and
expressions $P_1$--$P_4$ of Tab.~\ref{xtab:exp_pattern_expressions}). We
compared \NAME{}'s performance against state-of-the-art RE-constraint FSM
methods SMA and prefix-growth. We used the PRT dataset and obtained suitable RE
constraints from the PROSITE database~\cite{prosite}; the runtimes are shown in
log-scale in Fig.~\ref{fig:re-constraints}. We observed that DESQ was up to 2.5x
slower than SMA for $P_1$ and up to 1.3x slower than SMA on $P_2$.  We do not
give SMA results for $P_3$ and $P_4$ because the implementation produced
incorrect results (acknowledged by the original authors). We did not investigate
this further as the SMA source is not available. DESQ was roughly on par with
prefix-growth for $P_1$--$P_4$ (up to 1.3x) slower. The overhead of DESQ thus
appears acceptable.

\subsection{DESQ for Customized Subsequence Constraints}

We evaluated the performance of DESQ for customized subsequence constraints that
may arise in applications.

\textbf{Constraints.} We considered pattern
expressions that express constraints in information extraction (IE), natural
language processing (NLP), and customer behavior mining applications. These
expressions are shown in Tab.~\ref{xtab:exp_pattern_expressions}. Expressions
${N_1}$--${N_5}$ express constraints useful for IE and NLP applications and
are inspired from \cite{reverb, Google-ngrams, patty,finet}; these expressions were
used on the NYT dataset. Expressions ${A_1}$--${A_4}$ expresses constraints
useful for market-basket analysis and apply to AMZN.

\begin{figure}[t]
\centering
\includegraphics[scale=0.8]{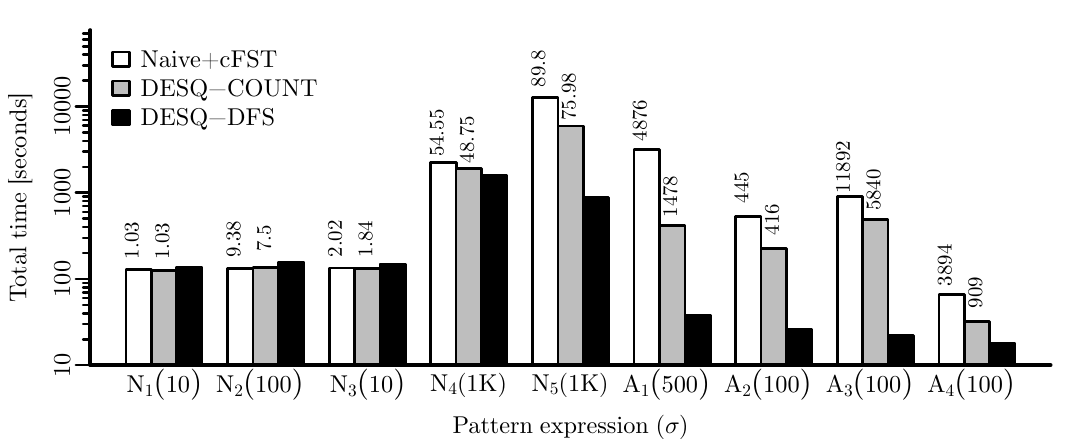}
\caption{Performance of DESQ mining algorithms. The numbers on top of the bars
indicate the average number of $P$-subsequences per input sequence.}
\label{fig:desq_mining_alg}
\end{figure}

\textbf{DESQ algorithms.} We evaluated the performance of
Na\"ive, \NAME-COUNT and \NAME-DFS on pattern expressions ${N_1}$--${N_5}$ and
${A_1}$--${A_4}$. The results are shown in Fig.~\ref{fig:desq_mining_alg}, which
also gives the minimum support threshold $\sigma$ used for each pattern
expression (chosen empirically). The runtimes are given in log-scale.  On the
NYT dataset, for expressions ${N_1}$--${N_3}$, \NAME-COUNT and \NAME-DFS had
similar performance and finished in a few minutes. For ${N_4}$--${N_5}$,
however, runtimes were higher and \NAME-DFS was significantly faster than
\NAME-COUNT (up to 14x). To gain insight into these results, we computed the
average number $\mu$ of $P$-sequences (average of $\sz{G_P^F(T)}$).\footnote{We
  averaged over input sequences $T$ for which $G_P^F(T) \neq \emptyset$.} These
numbers are shown above the bars for each pattern expression. We observed that
for small values of $\mu$, \NAME-COUNT and \NAME-DFS had similar performance,
whereas for larger values of $\mu$, \NAME-DFS was much more efficient. When
$\mu$ is small, the simple counting method of \NAME-COUNT is expected to work
well because few sequences are generated. The advanced pruning methods of
\NAME-DFS are then not needed.  When $\mu$ is large, however, \NAME-COUNT can
enumerate many sequences that turn out to be infrequent, which is
expensive. \NAME-DFS prunes many of these sequences early on and is thus more
efficient.

On the AMZN dataset (expressions $A_1$--$A_4$) \NAME-DFS consistently
outperformed \NAME-COUNT (up to 22x). This behavior is explained by the
observation that $\mu$ was large for all pattern expressions.

Based on these results, we conclude that \NAME-DFS consistently worked well in
our experiments. Although \NAME-COUNT was slightly faster in some cases, it
blew up on others. Thus we consider it generally safer to use \NAME-DFS in
practice.

\subsection{FST Optimizations}

\begin{table}[t]
\setlength{\tabcolsep}{3.5pt}
\centering
\caption{Runtimes of Na\"ive with cFST and openFST}
\begin{tabular}{llllllllll}
\toprule
 & $N_1$ & $N_2$ & $N_3$ & $N_4$ & $N_5$  & $A_1$  & $A_2$  & $A_3$  & $A_4$  \\
\midrule
OpenFST & 1h & 1h & 1hr & 6h  & $>$12h & $>$12h & $>$12h & $>$12h & 50m  \\
cFST    & 2m & 2m & 2m  & 30m & 3.5h   & 1h     & 10m    & 15m    & 1m   \\
\bottomrule
\end{tabular}
\label{tab:fst_optimizations}
\end{table}

We also compared cFST simulation with the state-of-the-art OpenFST library. We
used the Na\"ive ``generate-and-count'' approach on pattern expressions
$N_1$--$N_5$ and $A_1$--$A_4$. The results are shown in
Tab.~\ref{tab:fst_optimizations}. We observed that Na\"ive was orders of
magnitude faster when used with cFST simulation than when used with OpenFST.
This is because pattern expressions often translate to excessively large FSTs,
which are inefficient to simulate (see Tab.~\ref{tab:item_expressions} and
discussion on cFSTs in Sec.~\ref{sec:pattern_matching}). Moreover, OpenFST
cannot directly handle hierarchies and, as discussed in
Sec.~\ref{sec:pattern_matching}, and many of our pattern expressions cannot be
determinized. We conclude that cFST compression is effective.


\section{Related Work} 
\label{sec:related_work}

\textbf{Sequential pattern mining.} The problem of mining frequent sequential
patterns was introduced by Agarwal and Srikant~\cite{apriori}. Their Apriori
algorithm follows a candidate-generation-and-test approach. The subsequent GSP
algorithm~\cite{gsp} exploits the antimonotonicity property of sequential
patterns to efficiently generate and prune candidate sequences.
SPADE~\cite{spade} by Zaki also generates and prunes candidates, but it
operates on an inverted index structure representation of the database. Pei et
al. proposed the PrefixSpan algorithm~\cite{prefixspan}, which is based on a
more efficient pattern-growth approach that recursively grows frequent
prefixes using database projections.  \NAME-DFS can be seen as a
generalization of PrefixSpan to support arbitrary pattern expressions.
SPAM~\cite{spam}, which is similar to SPADE, uses an internal bitmap structure
for database representation and employs a pattern-growth approach to mine
frequent sequential patterns. A comprehensive discussion of these methods is
given in~\cite{Mabroukeh}.

\textbf{Subsequence constraints.}. There are many extensions to the
basic sequential pattern mining framework for supporting subsequence
constraints. GSP~\cite{gsp} and LASH~\cite{lash}, for example, allow gap
constraints and incorporate item hierarchies.  cSPADE~\cite{cspade} handles
length, gap and item constraints. Wu et al.~\cite{wu2014mining} consider
subsequences with periodic wild card gaps, i.e., subsequences where
consecutive items are separated by the same gap in the input. RE constraints
have been studied by~\cite{spirit, hackle_tree, prefix_monotone, petri_net};
these methods do not support capture groups. Some of the above constraints
(e.g., gap constraints) target the input sequence, whereas others (e.g.,
length constraints, RE constraints) target subsequences. Pattern expressions
unify both targets and allows us to express all of the above subsequence
constraints (e.g., see Tab.~\ref{tab:existing_relevance}) as well as
customized subsequence constraints that arise in FSM applications (e.g., see
Tab.~\ref{xtab:exp_pattern_expressions}).

\textbf{Pattern matching.} Our work is also related to to pattern
matching. There are many languages and systems for pattern matching over
sequences; e.g., for information extraction~\cite{SystemT, CPSL}, computational
linguistics~\cite{CQL}, complex event processing~\cite{dejavu}, and sequence
databases~\cite{Seshadri:1996,s-olap}. Our pattern expressions are simpler than
most pattern matching languages, yet expressive enough to specify many
subsequence constraints that arise in applications. Nevertheless, pattern
matching languages can conceivably be used to specify subsequence predicates and
mine $P$-frequent sequences using Na\"ive, i.e., by first enumerating all
matches and subsequently counting frequencies. Our experiments indicate that
this approach is infeasible for many subsequence constraints. Instead, it is
beneficial to integrate pattern matching and mining, e.g., along the lines of
\NAME-COUNT and \NAME-DFS. An interesting direction for future work is to
investigate to what extent such integration is possible for more powerful
pattern matching languages.


\section{Conclusions} \label{sec:conclusions}

In this paper, we introduced subsequence predicates as a general model for
unifying and extending subsequence constraints for FSM. We proposed pattern
expressions as a simple, intuitive way to express subsequence constraints,
suggested compressed finite state transducers as an underlying computation
model, and proposed the DESQ-COUNT and DESQ-DFS algorithms for efficient
mining. Our experiments indicate that DESQ is an efficient, general-purpose FSM
framework for common as well as customized subsequence constraints.




\bibliographystyle{IEEEtran}
\bibliography{IEEEabrv,references}
\end{document}